\documentclass[trackchanges,twocolumn]{aastex701}

\usepackage{amsmath}

\newcommand{\RLC}{R_{\mathrm{LC}}}
\newcommand{\PKHY}[1]{{{#1}}}
\submitjournal{ApJ}

\begin{document}

\title{A common four-beam geometry reveals altitude-stratified GeV pulses in canonical young pulsars}

\author[orcid=0000-0003-3476-022X,sname='Yeung']{Paul K. H. Yeung}
\affiliation{Institute for Cosmic Ray Research, University of Tokyo, 5-1-5 Kashiwa-no-ha, Kashiwa, Chiba 277-8582, Japan}
\email[show]{pkh91yg@icrr.u-tokyo.ac.jp}  

\author[orcid=0000-0001-6201-3761,sname='Saito']{Takayuki Saito} 
\affiliation{Institute for Cosmic Ray Research, University of Tokyo, 5-1-5 Kashiwa-no-ha, Kashiwa, Chiba 277-8582, Japan}
\email{}

\begin{abstract}

Despite the diversity and energy dependence of $\gamma$-ray pulse morphologies in Crab, Vela and Dragonfly, the phaseograms of these three canonical young pulsars can be organised within a single four-beam geometric template. Using \textit{Fermi} Large Area Telescope data, we fit the 60~MeV--3~GeV phaseograms with a mechanism-agnostic, geometry-first parametric model that incorporates phase-dependent Doppler shifts and constrains the three-dimensional locations and bulk motions of four emission sites. In each pulsar, the phaseogram admits a decomposition into two altitude-separated beam pairs.  The lower-altitude pair is produced by plasma with bulk motion close to azimuthal corotation, sharpening the main peaks. The higher-altitude pair shows a radially outward bulk-motion component, suggestive of inertial effects in a toroidally dominated magnetic field, and contributes bridge/shoulder emission and ripple-like modulations overlapping the main peaks. \PKHY{As a posteriori,} the lower-altitude pair is consistent with curvature-dominated outer-magnetospheric emission, while the higher-altitude pair is consistent with synchrotron-dominated emission from a current-sheet-like outflow. Higher-altitude site heights \PKHY{vary} from $\simeq 0.7$ (Crab, $\approx 1$~kyr) to $\simeq 1.1$--$1.4$ light-cylinder radii (Vela and Dragonfly, $\approx 10$~kyr). This unified four-beam, observation-driven geometry maps an altitude-dependent azimuthal tilt of pulsed $\gamma$-ray emission, providing an observationally anchored framework amenable to systematic tests and readily extensible to other young pulsars.

\end{abstract}



\section{Introduction}

Pulsars are among the most powerful laboratories for relativistic plasma physics. Their rapidly rotating, highly magnetised neutron stars sustain magnetospheres in which charged particles are accelerated to ultra-relativistic energies and radiate across the electromagnetic spectrum \citep{Michel1973}. Outside the region of near-rigid co-rotation, the outflow is expected to form a relativistic wind that feeds the surrounding pulsar wind nebula (PWN). Theories of global magnetohydrodynamic (MHD) and particle-in-cell (PIC) simulations emphasise the light-cylinder region, at cylindrical radius $\RLC$, as the approximate site where the magnetosphere transitions into a toroidally dominated wind and an equatorial current sheet develops \citep[e.g.][]{Goldreich1969,Coroniti1990,Bogovalov1999,Contopoulos1999,Spitkovsky2006,Tchekhovskoy2013,Kalapotharakos2014,Philippov2015}. Yet the three-dimensional structure of this transition zone, and its evolution with pulsar age and magnetisation, remain poorly constrained by observations.

High-energy observations with Fermi Large Area Telescope (LAT) have revealed a rich phenomenology of $\gamma$-ray phaseograms for young pulsars, typically showing two bright peaks and  bridge-like components between them \citep[e.g.][]{Smith2023,Abdo2010_VelaPSR,Leung2014,HESS_2023_VelaPSR,Kargaltsev2024,Abdo_Crab_2010,Yeung_CrabPSR_2020,CTAO-LST_CrabPSR_2024,Wang_thesis_2025,Lange2025}. These profiles are usually modelled with assumed particular acceleration sites and
radiation mechanisms---such as outer-gap, slot-gap or current-sheet models \citep[e.g.][]{Cheng1986,Arons1979,Ruderman1975,Daugherty1996,Romani1995,Dyks2003,Bai2010,Watters2009,Harding2021}. While such approaches have yielded important insights, they also introduce degeneracies: different microscopic prescriptions could simulate similar pulse shapes. A complementary strategy is to treat the observed $\gamma$-ray pulses primarily as geometric signatures of Doppler-boosted beams tied to bulk plasma flows, and to infer the underlying three-dimensional beam configuration directly from the phaseograms, with minimal assumptions about the emission mechanism \citep{Yeung2025}. In such a geometry-first strategy, we discuss mechanism-specific interpretations  only a posteriori, in an attempt to avoid circular reasoning.

In \cite{Yeung2025} we showed that a mechanism-independent two-beam model with purely tangential bulk flows
reproduces the gross GeV pulse morphology. However, fitting bridge/shoulder features with only two components required
highly flexible energy-dependent beam shapes, and the tangential constraint restricted the model’s predictive leverage on outflow-related Doppler
effects. We therefore introduce two additional beams and fit the bulk-flow directions explicitly, yielding a more
robust four-beam decomposition \PKHY{under the assumption of \emph{local} north-south symmetry (i.e. two pairs of beams, with each pair consisting of two identical antiparallel beams). In view of the increasing degeneracy expected when introducing additional components, we avoid higher-order decompositions.} 

Our localisation
of each emission site builds on the conventional light-cylinder-based view and recasts it as a two-dimensional $(r,z)$
description, providing a common geometric language for comparison with simulations and PWN-based spin-axis constraints
\citep[e.g.][]{Gaensler2006,Ng2004,Ng2008,Kirk2009,Porth2013,Kargaltsev2024}. \PKHY{Another noteworthy feature of our modelling framework is that the outflow-related $\gamma$-ray pulsation involves the continuous replacement of photon-emitting particles (Appendix~\ref{apx:virtual_corot}). This elucidates how an emission site outside the light cylinder could still operate even without requiring rigid co-rotation or superluminal material motion.}

\section{Sampling of LAT phaseograms}

We apply this framework to the Fermi-LAT data of three canonical young pulsars: Crab (PSR J0534+2200),
Vela (PSR J0835-4510) and Dragonfly (PSR J2021+3651). They span characteristic ages from very young ($\approx 1$~kyr) to adolescent ($\approx 10$~kyr) \citep{Albert_2021_Dragonfly, Smith2023, Manchester_2005_ATNF}. We construct four phaseograms per pulsar, for these divided energy segments: 0.06--0.15, 0.15--0.36, 0.36--0.9 and 0.9--3~{GeV}. At these energies, curvature and synchrotron emission treated with a single Doppler transformation are expected to dominate. We do not assign per-photon probability weights. Instead, all unpulsed and background contributions are absorbed into a constant background term $C_{\mathrm{bkg}}$ fitted simultaneously with the pulsed components $C_i(\psi,\varepsilon)$.

For Crab and Vela pulsars we use Pass~8 LAT data and select SOURCE-class events accumulated over 17~yr of observations, within a circular region of interest (a $3^\circ$ radius) centred on the pulsar position. We further filter the data by accepting only the good time intervals where the region of interest was observed at a zenith angle less than $90^{\circ}$ in order to suppress the contamination from the albedo of Earth, following the LAT collaboration recommendations \citep{Atwood2009}. Photon arrival times are transformed to the Solar system barycentre, and observed phases are computed with the \textsc{PINT} timing package using dedicated or catalogued ephemerides appropriate for each pulsar. Ephemerides of the Crab pulsar are provided by Jodrell Bank, while an optimised ephemeris of the Vela pulsar is provided by Matthew Kerr. We reconstruct the phaseograms of Dragonfly from an event list in the database of Fermi 3rd Pulsar Catalog \citep{Smith2023}, \PKHY{where pulse phases had already been computed}. 

\PKHY{For Crab, Vela and Dragonfly, we divide the full phase into 250, 500 and 80 bins respectively to obtain the baseline phaseograms (Fig.~\ref{fig:phaseograms}). We have also confirmed the robustness of our fittings by using alternative bin sizes of 0.75--1.25 times the baseline one for each pulsar.}

\section{Mechanism-independent four-beam geometry}

Our modeling framework parametrises only the macroscopic emission geometry---the inclination angle $\Theta_A$ of the spin axis from the line of sight\footnote{\PKHY{We denote it by $\Theta_A$ so as to distinguish it from the PWN-based estimate $\zeta$.}}, three-dimensional locations of emission sites, velocities of bulk motion and orientations of the beam axes---and infer these directly from the phaseograms.

Compared to our earlier two-beam  models \citep{Yeung2025}, the present four-beam scheme trades some flexibility for a simpler and more unified geometry. Although we now describe four beams and explicitly parametrise the bulk-flow directions, all geometric parameters are enforced to be common across four simultaneously fitted phaseograms. We also neglect the extension of each emission site  and replace the structured count-rate function  with an economical beam-shape prescription. 

We assume that the GeV emission can be decomposed
into four components: two in the northern hemisphere
and two in the southern hemisphere. \PKHY{Each pair is
described by a shared set of geometric, kinematic and beam-shape parameters:}
the second site is obtained from the first by a
$180^\circ$ rotation in the meridional plane (point
symmetry about the neutron star), so that north–south
symmetry is enforced.
This symmetry is imposed \emph{locally}, only on the spatially unresolved emission geometry ($r\lesssim R_{\rm LC}$) that shapes the $\gamma$-ray phaseograms. 

\subsection{Bulk motion and phase-dependent Doppler boosting}

At each emission site, the radiating plasma is assumed to move with a bulk velocity $\bm{v}$ of magnitude $v$, described by a Lorentz factor and two orientation angles. For one beam pair we denote the bulk Lorentz factor by $\gamma_\mathrm{Lor}$ and its orientation by zenith and azimuthal angles $(\theta_M, \theta_N)$ measured from the spin axis and a fixed meridional plane, respectively. Generalising equation (5) of \cite{Yeung2025}, we obtain an expression for the inverse Doppler factor (i.e. the ratio of the plasma-frame energy $E_\mathrm{bulk}$ to the detected energy $E_\mathrm{det}$) for each beam: 
\begin{equation}
\left\{
\begin{aligned}
\varepsilon &= \frac{E_\mathrm{bulk}}{E_\mathrm{det}}
             = \frac{1}{\gamma_{\mathrm{Lor}}\left(1 - \frac{v}{c} \cos\psi_{\mathrm{vel}}\right)}\\[4pt]
\cos\psi_{\mathrm{vel}}
&= -\,\sin\Theta_{\mathrm{A}}\,
     \cos\bigl(\theta_{\mathrm{N}} + 2\pi \varphi)\,
     \sin\theta_{\mathrm{M}} \\
&\quad + \cos\Theta_{\mathrm{A}}\,
        \cos\theta_{\mathrm{M}} \, 
\end{aligned}
\right. ,
\label{eqn:Doppler}
\end{equation}
where $\gamma_{\mathrm{Lor}}$ is the corresponding Lorentz factor, $\psi_{\mathrm{vel}}$ is the angle between $\bm{v}$ and the line of sight, $c$ is the speed of light, and $\varphi$ denotes the pulsar’s rotational angle. 

Even though individual particles cool and the bulk flow undergoes gradual acceleration or deceleration, the fitted parameters $(\gamma_{\rm Lor},\theta_M,\theta_N)$ can  be interpreted as effective, emission-weighted averages over the pulse-emitting segment of the flow. Within this standard approximation, the “plasma frame’’ is understood as the momentarily comoving inertial frame of a representative plasma element along each beam, while the ``corotating frame" is used only as a non-inertial geometric coordinate system for defining the orientation of the emission pattern. Such a segregation between frames is further clarified in Appendix~\ref{apx:virtual_corot}.

\subsection{Simulation of the detected count rate}

We make a minimal assumption of an axisymmetric angular emissivity kernel (i.e. an intrinsically circular beam ‘core’) for each component beam in the plasma frame, enabling a stable component decomposition. Within our framework, we formulate the detector-frame differential photon count rate contributed by beam $i$  as a product of an intrinsic, plasma-frame angular pattern and an energy-dependent scaling factor,
\begin{equation}
C_i(\psi,\varepsilon)
= A_i\,\varepsilon^{\eta_i}
\exp\!\left[
  -\left(\frac{\psi}{\Psi_{c,i}}\right)^{\beta_i}
\right],
\label{eq:Ci_GND}
\end{equation}
where $\psi$ is the angle, measured in the co-moving (plasma) frame, between the line of sight and the beam axis, and $\varepsilon \equiv E_{\mathrm{bulk}}/E_{\mathrm{det}}$ is the inverse Doppler factor defined in Equation~(\ref{eqn:Doppler}). The exponential factor represents the baseline, energy-averaged generalised normal distribution (GND) angular profile (‘kernel’) of the beam in the plasma frame, while the factor $\varepsilon^{\eta_i}$ maps this pattern into the detector frame. \PKHY{$A_i$ sets the overall normalisation, $\eta_i$ controls the power-law dependence on $\varepsilon$, $\Psi_{c,i}$ is the characteristic angular width of the GND profile, and $\beta_i$ is its angular sharpness index.}

For a given pulsar and a given energy band, the model-predicted number of counts in phase bin $k$ is
\begin{equation}
\lambda_k = C_{\mathrm{bkg}} + \sum_{i=1}^4 C_i(\psi,\varepsilon).
\end{equation}

\begin{figure*}[t]
    \centering
    \includegraphics[width=.32\textwidth]{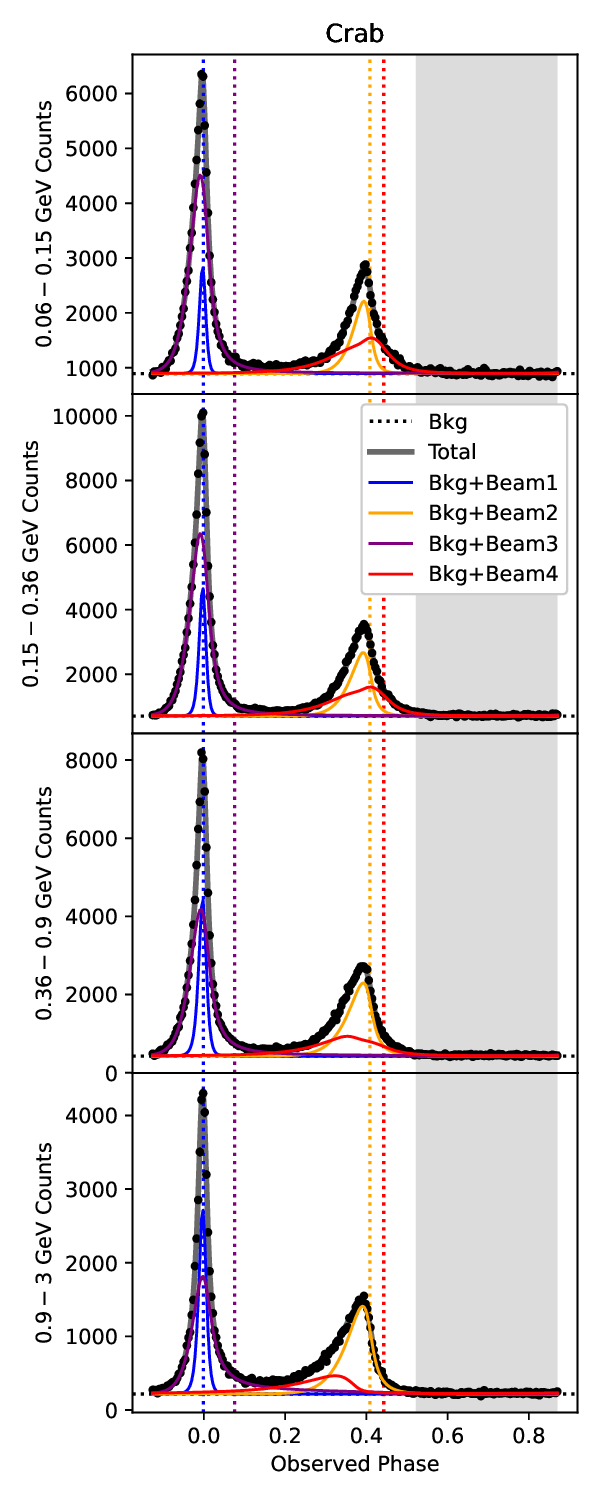}
    \includegraphics[width=.32\textwidth]{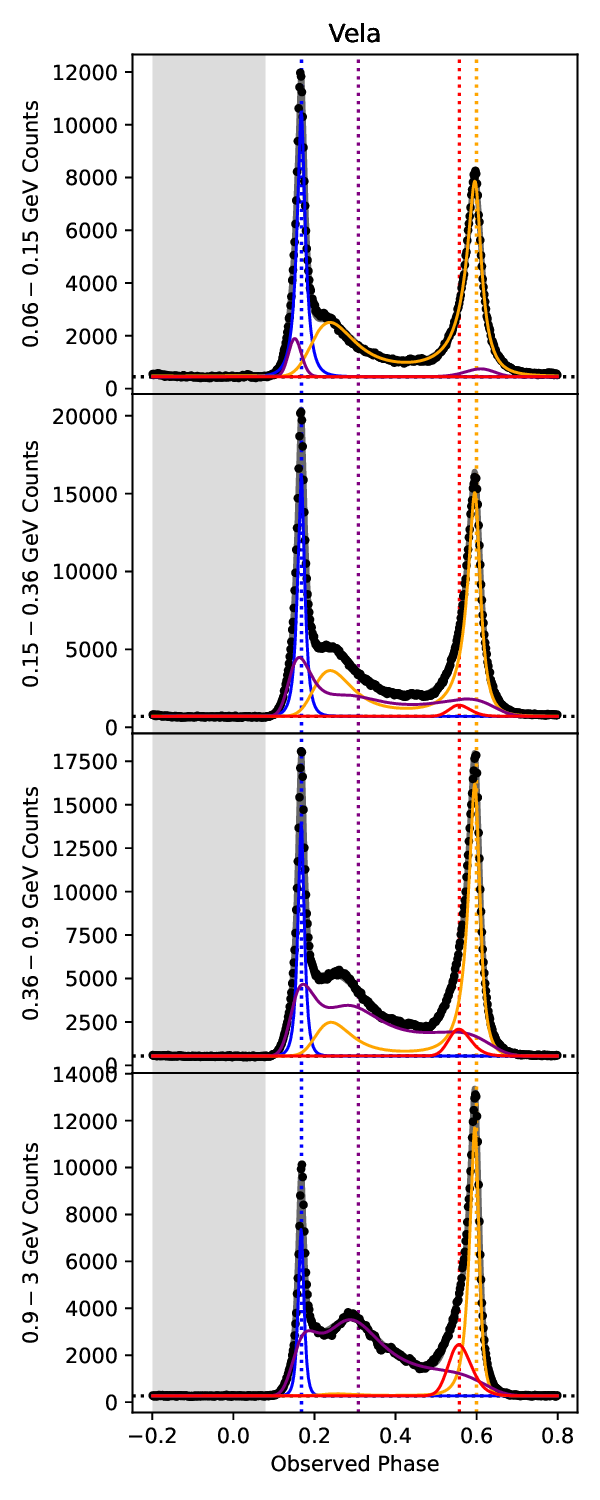}
    \includegraphics[width=.32\textwidth]{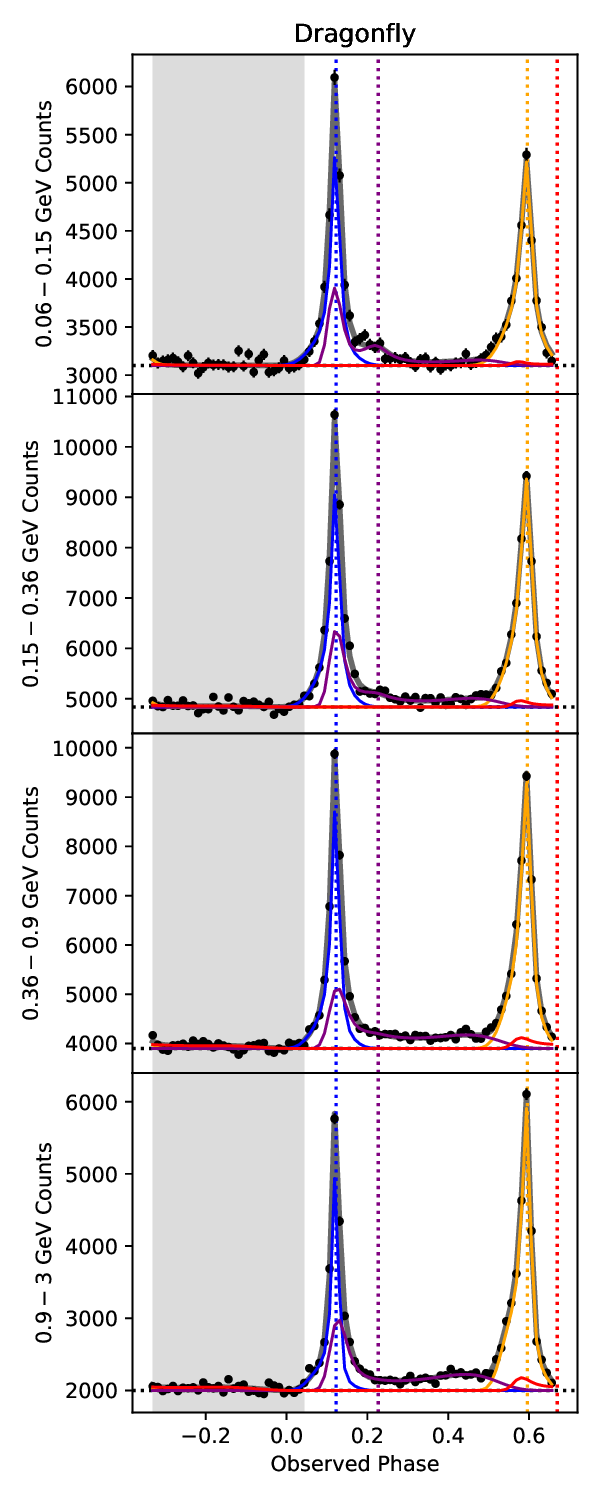}
    \caption{Phaseograms and four-beam decomposition for Crab, Vela and Dragonfly. Each panel shows the observed Fermi-LAT phaseogram in four energy bands between 60~MeV and 3~GeV, together with the best-fitting four-beam model and the contributions from the individual beams. The  lower-altitude beam pair (Beams~1 and~2) reproduces the main double-peaked structure, while the higher-altitude beam pair (Beams~3 and~4) accounts for the bridge-like and inter-pulse emission. Each vertical dotted line spanning over a column of panels indicates the pulse phase when a beam axis is closest to our l.o.s. (not to be confused with the actually observed peak phase, which is further modulated by the Doppler shift). The grey shaded intervals were defined as the off-pulse phase range in previous literature \citep{Aleksic_Gap_2012, HESS_Vela_2018, Wang_thesis_2025}.}
    \label{fig:phaseograms}
\end{figure*}

\subsection{\PKHY{$\eta_i$ as an effective Doppler-response index}}\label{main:eta_i}

Although $\eta_i$ formally modulates the amplitude $A_i$ through the factor $\varepsilon^{\eta_i}$, in practice it is best viewed as an \emph{effective Doppler-response index} that parametrises how the band-limited intensity of beam $i$ reacts to phase-dependent reweighting of the contributing $E_{\rm bulk}$ distribution through $\varepsilon(\varphi)$.
In this role, $\eta_i$ can absorb residual energy dependence of the intrinsic beam-shape parameters (such as $\Psi_{c,i}$ and $\beta_i$) that are not explicitly modelled, as well as other unresolved complexities  that project onto the observed modulation within a finite $E_{\rm det}$ band (Appendix~\ref{apx:crosscheck}).
We therefore treat $\eta_i$ as a phenomenological, band-integrated parameter controlling the overall response of beam $i$ to $\varepsilon$ across 60~MeV--3~GeV, by construction distinct from the conventional photon index.

Within this interpretation, and given the typically mildly relativistic bulk Lorentz factors $\gamma_{\mathrm{Lor}}$ inferred for most beams, even apparently large values such as $\eta_i\sim10^3$ in Table~\ref{tab:beamshape_all} primarily indicate a strong sensitivity of the fitted band-limited intensity to small fractional changes in $\varepsilon$, rather than requiring extreme intrinsic spectral slopes. This motivates treating $\eta_i$ as a compact, model-agnostic summary of the band-integrated Doppler response, which can be confronted with physical prescriptions through forward-folding. \PKHY{Further interpretation of $\eta_i$ is presented in Appendix~\ref{apx:eta_i}.} The correlation between $\eta_i$ and the bulk Lorentz factors, and its implications for interpreting apparently large $\eta_i$, are discussed in the Appendix~\ref{apx:crosscheck}.

\section{Results and Discussion}

\begin{figure*}[t]
    \centering
    \includegraphics[width=.32\linewidth]{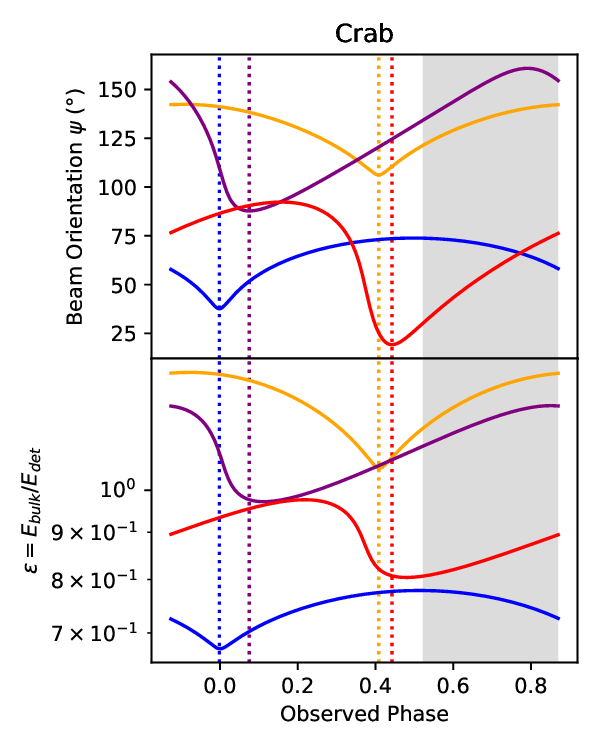}
    \includegraphics[width=.32\textwidth]{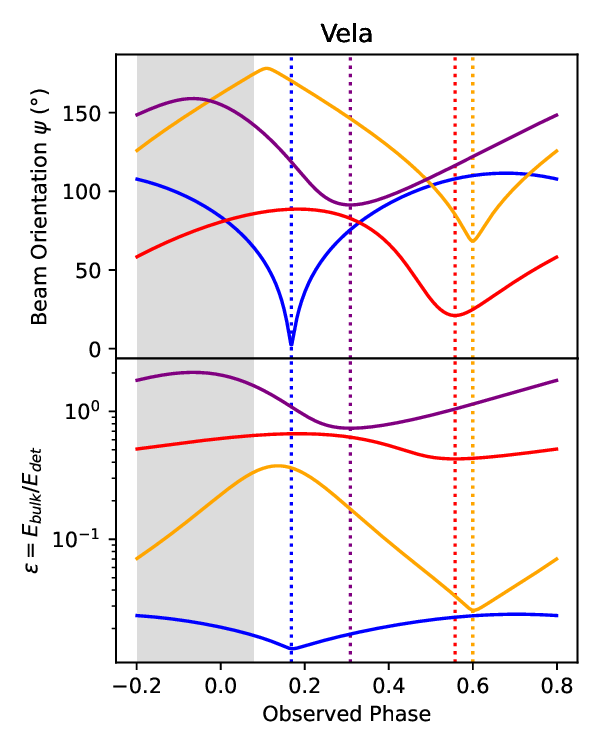}
    \includegraphics[width=.32\textwidth]{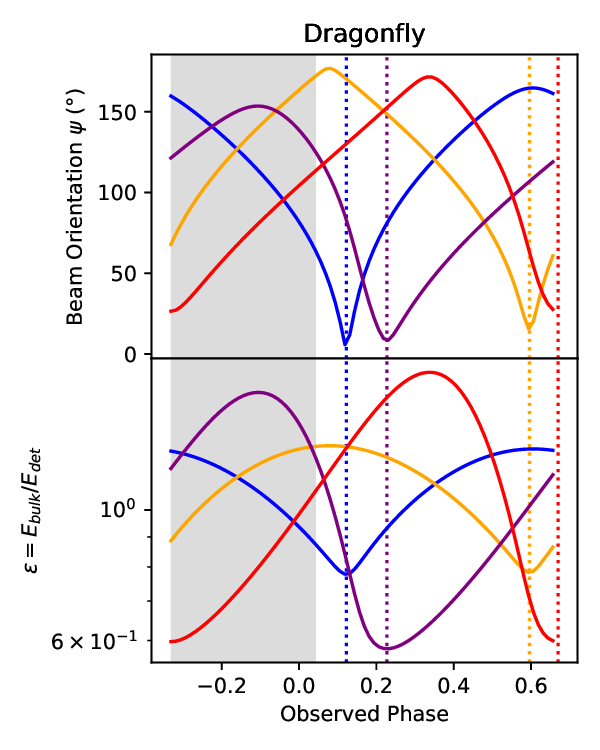}
    \caption{Distributions of $\psi$ (beam orientation) and $\varepsilon$ (inverse Doppler factor) for the Crab, Vela, and Dragonfly pulsars. Each column corresponds to one pulsar, with the top row showing the orientation-angle distribution $\psi$ and the bottom row showing the Doppler parameter $\varepsilon$. Both quantities are energy-independent and were derived from the same four-beam fits used in Table~\ref{tab:geometry}. The colours and the dotted line-style have the same meanings as in Fig.~\ref{fig:phaseograms}.}
    \label{fig:psi_epsilon}
\end{figure*}

\begin{figure*}[t]
    \centering
    \includegraphics[width=.96\textwidth]{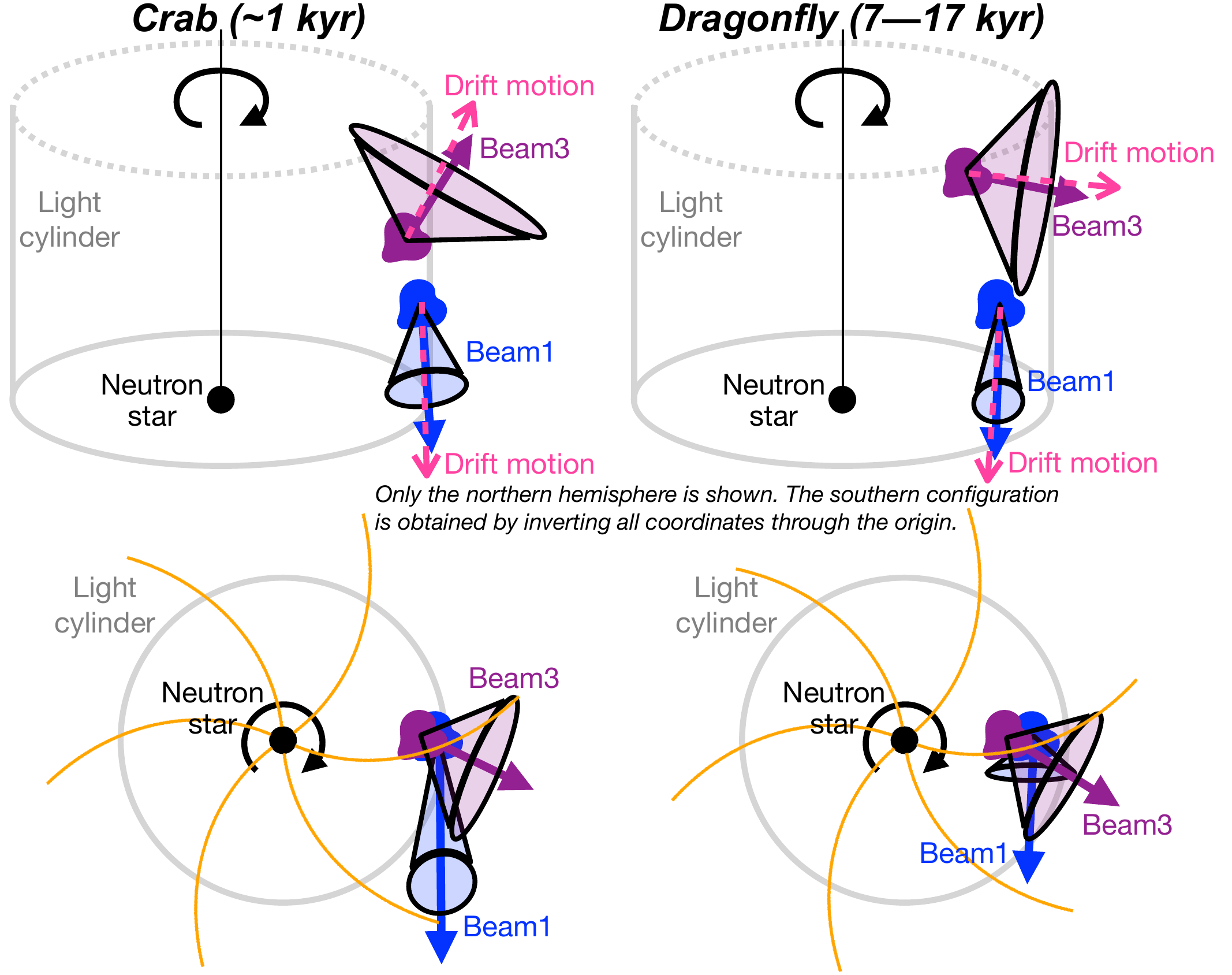}
    \caption{Resolved four-beam templates for the very young Crab and adolescent Dragonfly pulsars. Schematic representation of the four-beam configuration for each pulsar, showing the locations of the lower-altitude and higher-altitude emission sites relative to the light cylinder. Only the northern hemisphere is shown, while the southern configuration is obtained by a point reflection through the origin, i.e. by flipping the signs of all coordinates. The inferred beam axes are interpreted as approximately tangential to the local magnetic-field-line directions at the fitted emission sites. In particular, the Beam1 axis of the lower-altitude site lies close to the last closed field lines near the light-cylinder boundary and has only a small radial component in its inferred beam-axis direction. Whereas, the Beam3 axis of the higher-altitude site is expected to be shaped primarily by inertial sweepback  \citep[the toroidal, Parker-spiral-like  magnetic field overlaid in each top view;][]{Parker1958,Michel1973,Bogovalov1999,Kirk2009} rather than uniquely tracing the dipole direction. For clarity, the opening angle of each cone is schematic and does not represent the fitted angular width or sharpness of that beam, which is energy-dependent.}
    \label{fig:schematic}
\end{figure*}

The four-beam framework provides a concise description of the observed phaseograms (Fig.~\ref{fig:phaseograms}), enabling a
controlled geometric comparison among pulsars. The model-predicted distributions of the instantaneous beam orientation and plasma-frame energy are shown in Fig.~\ref{fig:psi_epsilon}. The inferred geometries and bulk-flow directions are broadly
consistent across our cross-checking variants (Appendix~\ref{apx:crosscheck}). Based on the four-beam decomposition, we propose a common pattern in how the $\gamma$-ray emission of young pulsars is distributed between azimuthally corotation-like and radially outflow-tilted components in an altitude-stratified manner. Schematic views of the inferred geometric configurations are shown in Fig.~\ref{fig:schematic}. The uncertainties quoted in Table~\ref{tab:geometry} are the quadratic sum of the statistical errors and the systematic component estimated from the $E_{\max}$ band-edge checks (Appendix~\ref{apx:band_edge}). 

The polar axis of the PWN's equatorial torus was constrained to be
inclined from our line of sight by $\zeta=61^{\circ}$--$63^{\circ}$, $63.6^{\circ}$ and
$85^{\circ}$--$88^{\circ}$ for Crab, Vela and Dragonfly, respectively
\citep[e.g.][]{Ng2004,VanEtten2008,Jin2023}. In our four-beam fits we obtain
spin-axis inclinations of $\Theta_{\mathrm{A}} = 55.7 \pm 5.6^{\circ}$,
$54.9^{\circ} \pm 0.6^{\circ}$ and $80.9^{\circ} \pm 1.4^{\circ}$ for these  pulsars
(Table~\ref{tab:geometry}). Our fitted $\Theta_{\mathrm{A}}$ values remain compatible with the
PWN-based constraints.

Beams~1 and~2 form a lower-altitude pair whose emission sites remain radially close to the light-cylinder boundary in all three pulsars, with cylindrical radii $r \simeq 0.7$--$1.1\,\RLC$ and modest altitudes above the equatorial plane. These beams naturally account for the two main $\gamma$-ray peaks in the phaseograms and can be interpreted as Doppler-boosted emission from bulk plasma flows whose directions are characterised, for the lower-altitude pair, by the zenith and azimuthal angles $(\theta_M,\theta_N)$ listed in Table~\ref{tab:geometry}. In our fits, $\theta_N \simeq 80^{\circ}$--$100^{\circ}$ for all three
pulsars, consistent with bulk motions that remain nearly tangential and
closely tied to quasi-rigidly co-rotating field-line bundles near the last
closed field lines. \PKHY{Within this small sample,} the decline in $\theta_M$ with age (from
$\simeq 166^{\circ}$ in the Crab to $\simeq 147^{\circ}$ in Vela and
$\simeq 97^{\circ}$ in Dragonfly) \PKHY{tentatively} suggests that, if flows roughly follow
field lines, their tangents in older systems progressively align with the
equatorial plane.

The relatively large bulk Lorentz factor inferred for the lower-altitude pair in Vela,
$\gamma_{\mathrm{Lor},1} \simeq 40$, could be interpreted as an \emph{effective}
measure of particularly fast flow segments (or intermittently faster streams) in the outer magnetosphere required to reproduce the observed Doppler-shift modulation, rather than implying that the entire region shares a single bulk Lorentz factor. Because the model likelihood becomes progressively less sensitive once $\gamma_{\rm Lor,1}$ enters the strongly relativistic regime, our Vela fit primarily requires $\gamma_{\rm Lor,1}\gtrsim 10$ rather than constraining its precise value. By contrast, the crosschecks yield a conservative upper limit $\gamma_{\rm Lor,1}\lesssim 1.5$ for Crab and Dragonfly. Relevant details are provided in the Appendix~\ref{apx:crosscheck}.

In our fits for Vela, these beams are anchored almost on the light-cylinder surface in cylindrical radius ($d\sin\theta_{\rm B} \simeq 1.0 R_{\rm LC}$), along field-line bundles that remain approximately corotation-like in azimuth near the last closed field lines, but the plasma
there is already highly relativistic. Independent very-high-energy observations support this picture:
the detection of pulsed emission up to $\gtrsim 20~\mathrm{TeV}$ in Vela implies
the  maximal Lorentz factor $\gamma_{\max} \gtrsim 4\times 10^{7}$ for individual electrons, and joint GeV–TeV spectral
modelling constrains the bulk Lorentz factor of the post-light-cylinder flow (at radii of a few $R_{\rm LC}$) to be of order \(\sim 5\)–\(10\) \citep{HESS_2023_VelaPSR}. Within this context, a larger
$\gamma_{\mathrm{Lor},1}$ for the 60~MeV–3~GeV lower-altitude beam pair in Vela indicates that the pulse-emitting plasma in Vela’s outer magnetosphere is more  relativistic, and that stronger Doppler beaming appears to be required in our fits to reproduce its sharp high-energy main peaks.

\subsection{Altitude-dependent azimuthal orientation of pulse emission}\label{WindZone}

By contrast, Beams~3 and~4 form a higher-altitude pair whose emission sites occupy smaller cylindrical radii and higher altitudes than the primaries. In the Crab pulsar, this pair resides at $r \simeq 0.9\,R_{\mathrm{LC}}$ and heights of $z \approx 0.7\,R_{\mathrm{LC}}$ above the equatorial plane,  in regions that are not yet cleanly detached from the outer magnetosphere. In Vela and Dragonfly, however, the higher-altitude beam pair is pushed vertically outward to representative heights of $z \approx 1.4$ and $z \approx 1.1\,R_{\mathrm{LC}}$, respectively, while their cylindrical radii decrease to $r \simeq 0.6$--$0.7\,R_{\mathrm{LC}}$, \PKHY{possibly suggesting} an age-dependent vertical inflation of the rotation-decoupled outflow
between the very young Crab and the two adolescent pulsars, as well as a geometric reorganisation of the
higher-altitude outflow. \PKHY{Nevertheless, these trends need to be confirmed with a larger sample of broadly aged pulsars.} Table~\ref{tab:geometry} further shows that the lower-altitude and higher-altitude emission sites remain nearly co-longitudinal, with azimuthal separations $|\Delta\varphi_{\rm site}|\lesssim12^{\circ}$, so their geometric contrast is dominated by their different heights and, secondarily, by their different cylindrical radii (see the Appendix~\ref{apx:crosscheck} for BIC tests against the co-location limit). 

At such high altitudes of the Beams~3--4 sites, the magnetic field is expected to be significantly wound up and toroidally dominated, and global MHD simulations predict that the magnetosphere relaxes into a wind and the striped-wind current sheet begins to develop, so that the plasma flows can no longer be treated as  co-rotating with the neutron star \citep[e.g.][]{Coroniti1990,Bogovalov1999,Spitkovsky2006,Tchekhovskoy2013,Kalapotharakos2014,Kirk2009,Porth2013}. In our framework, this magnetosphere--wind transition may correspond to  a modest increase in the bulk Lorentz factor for the higher-altitude beam pair in the older systems. 

The bulk Lorentz factors inferred for the higher-altitude pair,
$\gamma_{\mathrm{Lor},3} \simeq 1.03-1.43$ in all three pulsars, could 
be interpreted in the context of the extended geometry of Beams~3 and~4.
These components are associated with broad, sheet-like emission regions in the
outflow-transition layer, rather than with narrow,
jet-like structures. In such a configuration, mildly relativistic bulk
motions with $\gamma_{\mathrm{Lor},3} \lesssim 1.5$ are expected to provide 
adequate Doppler boosting to enhance the visibility of the higher-altitude outflow
while preserving the broad, bridge-like modulation. Moreover, the fitted
$\gamma_{\mathrm{Lor},3}$ should be regarded as an \emph{effective} Lorentz factor
along the beam direction: the true flow in the transition layer may well
contain ultra-relativistic components, but turbulence, pitch-angle spreads and
the extended current-sheet geometry can dilute the net beaming towards the
observer so that only a modest effective Lorentz factor is manifested in the
60~MeV--3~GeV pulse profiles.
Cross-checks in the Appendix~\ref{apx:crosscheck} suggest a conservative upper limit of $\gamma_{\rm Lor} \lesssim 1.5$ on the bulk Lorentz factor of the plasma responsible for the pulsed emission in Crab and Dragonfly.

The fitted geometries also hint at a mild collimation of the higher-altitude outflow. As the emission sites move to larger $|z|$, the inferred cylindrical radii $r$ tend to decrease, indicating that the higher-altitude outflows are gently focused toward the spin axis. This behaviour is qualitatively consistent with hoop stresses in the toroidal magnetic field, which naturally narrow the outflows as the plasma propagates outward \citep{Kennel1984, Begelman1994,Komissarov2003}. In the very young, highly magnetised Crab pulsar this collimation is partly masked by strong magnetic confinement that keeps the outflow zones still entangled with the outer magnetosphere, whereas in the less magnetised, adolescent Vela and Dragonfly the reduced field tension allows both vertical inflation of the emission zones and a clearer outflow-dominated geometry.

The azimuthal geometry of the higher-altitude beam pair further clarifies their dynamical state. In the limit of rigid corotation the bulk flow would be purely tangential, corresponding to $\phi_{\mathrm{M}} \simeq \phi_{\mathrm{N}} \simeq 90^{\circ}$. Instead, Table~\ref{tab:geometry} shows that the fitted $\phi_{\mathrm{N}}$ values for Beams~3 and~4 lie well below $90^{\circ}$ in all three pulsars: in the Crab they are consistent, within the uncertainties, with flow that has  essentially no tangential motion ($\phi_{\mathrm{N}} \simeq 0^{\circ}$), whereas in Vela and Dragonfly they cluster around $\phi_{\mathrm{N}} \simeq 30$--$40^{\circ}$. These angles imply that the plasma sampled by the higher-altitude beam pair already carries a substantial radial component and has slipped out of azimuthal corotation, beginning to stream radially outward. Across the three pulsars the beam axes of the lower-altitude and higher-altitude pairs are azimuthally separated by $\Delta\varphi_{\rm beam}\simeq 40^{\circ}$--$65^{\circ}$, implying that the outflow-tilted beams sweep past our line of sight substantially later in pulsar-frame time than the primaries, a behaviour naturally expected if the outflow-transition layer are swept back by plasma inertia and a growing toroidal (Parker-spiral-like) field component \citep{Parker1958,Michel1973,Bogovalov1999,Kirk2009}.

\begin{table}[t]
\small
\setlength{\tabcolsep}{2pt}
\centering
\caption{Geometric and kinematic parameters of the four-beam configurations. Quoted uncertainties are the quadratic sum of statistical and systematic errors from the band-edge tests described in the Appendix~\ref{apx:band_edge}.}
\label{tab:geometry}
\begin{tabular}{@{}lccc@{}}
\hline
     & Crab & Vela & Dragonfly \\
\hline
$\Theta_{\rm A}$                 ($^\circ$)        & $55.7 \pm 5.6$   & $54.9 \pm 0.6$   & $80.9 \pm 1.4$   \\
\hline
\multicolumn{4}{c}{Beams 1 and 2} \\
\hline
$d\sin\theta_{\rm B}/R_{\rm LC}$    & $0.96 \pm 0.13$  & $1.01 \pm 0.01$  & $0.74 \pm 0.01$  \\
$d\cos\theta_{\rm B}/R_{\rm LC}$    & $0.50 \pm 0.05$  & $0.37 \pm 0.01$  & $0.53 \pm 0.07$  \\
$t_0$                            (phase)           & $0.122 \pm 0.005$& $0.286 \pm 0.001$& $0.260 \pm 0.001$\\
$\gamma_{\mathrm{Lor},1}$                    & $1.12 \pm 0.07$& $37.3 \pm 0.4$   & $1.032 \pm 0.001$\\
$\theta_{\rm M}$                 ($^\circ$)        & $165.7 \pm 0.9$  & $146.9 \pm 0.4$  & $97.1 \pm 0.5$   \\
$\theta_{\rm N}$                 ($^\circ$)        & $87.3 \pm 2.7$   & $82.6 \pm 0.1$   & $94.3 \pm 0.7$   \\
$\theta_{\rm C}$                 ($^\circ$)        & $161.9 \pm 1.0$  & $123.3 \pm 0.6$  & $96.1 \pm 0.3$   \\
$\theta_{\rm Q}$                 ($^\circ$)        & $90.2 \pm 1.8$   & $87.9 \pm 0.1$   & $94.5 \pm 0.7$   \\
\hline
\multicolumn{4}{c}{Beams 3 and 4} \\
\hline
$D\sin\phi_{\rm B}/R_{\rm LC}$    & $0.92 \pm 0.08$  & $0.60 \pm 0.01$  & $0.64 \pm 0.02$  \\
$D\cos\phi_{\rm B}/R_{\rm LC}$    & $0.74 \pm 0.15$  & $1.37 \pm 0.01$  & $1.14 \pm 0.18$  \\
$T_0$                            (phase)           & $0.137 \pm 0.001$& $0.395 \pm 0.005$& $0.304 \pm 0.003$\\
$\gamma_{\mathrm{Lor},3}$                    & $1.027 \pm 0.048$& $1.43 \pm 0.01$  & $1.153 \pm 0.007$ \\
$\phi_{\rm M}$                   ($^\circ$)        & $34.9 \pm 1.5$   & $31.2 \pm 0.6$   & $105.3 \pm 5.5$  \\
$\phi_{\rm N}$                   ($^\circ$)        & $9.4 \pm 11.1$   & $36.8 \pm 0.6$   & $33.7 \pm 0.8$   \\
$\phi_{\rm C}$                   ($^\circ$)        & $36.5 \pm 1.4$   & $33.8 \pm 0.6$   & $107.4 \pm 5.4$  \\
$\phi_{\rm Q}$                   ($^\circ$)        & $26.9 \pm 9.3$   & $37.0 \pm 0.6$   & $34.0 \pm 0.8$   \\
\hline
\multicolumn{4}{c}{Additional geometrical properties$^{\mathrm{a}}$} \\
\hline
$\Delta\theta$$^{\mathrm{b}}$ ($^{\circ}$)    &     3.9 & 23.9 & 1.0      \\
$\Delta\phi$$^{\mathrm{b}}$ ($^{\circ}$)     &     10.3 & 2.6 & 2.1        \\
$\Delta\varphi_{\rm site}$$^{\mathrm{c}}$ ($^{\circ}$)   & $-4.4$ & $-12.4$ & $4.9$ \\
$\Delta\varphi_{\rm beam}$$^{\mathrm{c}}$ ($^{\circ}$)   & $58.9$ & $38.5$ & $65.4$ \\
\hline
\end{tabular}

\footnotesize{\textit{Notes.} $^{\mathrm{a}}$ These additional properties are computed by substituting some iterated parameters into specific equations. \\
$^{\mathrm{b}}$ Angular separation between a beam axis and a bulk flow direction ($\Delta\theta$ for Beams 1 and 2, $\Delta\phi$ for Beams 3 and 4). These are computed using the scalar product formula. \\
$^{\mathrm{c}}$ Azimuthal separation between the lower-altitude and higher-altitude emission sites ($\Delta\varphi_{\rm site}$), and azimuthal separation between the corresponding beam axes ($\Delta\varphi_{\rm beam}$). Relevant formulae are shown in Appendix~\ref{apx:az_sep}. }
\end{table}

As a whole, the fitted geometric and kinematic properties of Beams 3 and 4 across the three pulsars are consistent with the weakening of magnetic tension and the increasing dominance of toroidally structured, rotation-decoupled outflow, motivating a view of the outflow-transition layer as an extended $(r,z)$ surface rather than a purely light-cylinder-defined radius.
Our inferred age-dependent trend for their altitudes is consistent with a monotonic inflation of the outflow-transition layer across these three canonical young pulsars, motivating a simple age-ordered picture that can be tested with a larger sample in future work.

\subsection{Energy dependencies of beam morphologies}

Beyond the global, energy-averaged geometry, we quantify how the pulse morphology varies with energy within the same four-beam configuration. For each pulsar and each of the four 60~MeV--3~GeV phaseograms, we model the lower-altitude (Beams~1--2) and higher-altitude (Beams~3--4) pairs by a generalised normal distribution (GND) in the plasma-frame angle $\psi$ from the beam axis, and assign the amplitude a power-law dependence on the inverse Doppler factor $\varepsilon$. In this way, our framework incorporates effective dependence on the plasma-frame energy $E_\mathrm{bulk}$ within a detected energy $E_\mathrm{det}$ band, in addition to band-to-band variations. The fitted beam-shape parameters and the constant background level are listed for Crab, Vela and Dragonfly in Table~\ref{tab:beamshape_all}. The need for energy-dependent sharpness indices is tested via a BIC comparison in the Appendix~\ref{apx:crosscheck}.

The power-law indices $\eta_1$ and $\eta_3$ quantify the effective dependence of the beam intensities on  $E_\mathrm{bulk}$ (equivalently,  on  $\varepsilon$) within an $E_\mathrm{det}$ band. Their positive sign implies that, for our finite detector bands—largely below the spectral cutoffs of these pulsars \citep{Smith2023}—band-limited photon counts can be preferentially enhanced at viewing directions with larger $\varepsilon$. Their apparently large values therefore indicate strong sensitivity of the band-limited intensity to modest changes in $\varepsilon$. Further interpretation of $\eta_1$ and $\eta_3$ is provided in Appendix~\ref{apx:eta_i}.

Across all three pulsars, the lower-altitude pair shows a small to moderate band-to-band variation in its angular width $\Psi_{c,1}$, with little change in $\beta_1$. In Crab, $\Psi_{c,1}$ stays near $\approx2$--$3^{\circ}$ and $\beta_1$ remains $\approx1.35$ from 60~MeV to 3~GeV, consistent with a largely geometry-driven double-peaked structure. In Vela and Dragonfly, the lower-altitude pair is broader in angle but more Gaussian-like: $\Psi_{c,1}$ narrows from $\simeq19^{\circ}$ to $\simeq13^{\circ}$ in Vela and from $\simeq7^{\circ}$ to $\simeq5^{\circ}$ in Dragonfly, with $\beta_1\approx2$.

By contrast, the higher-altitude pair is broader in angle and, in Crab, varies more strongly with energy, in line with its role in producing bridge/shoulder emission. In Crab, $\Psi_{c,3}$ increases from $\approx10^{\circ}$ at 60--150~MeV to $\approx30^{\circ}$ at 0.9--3~GeV, with an accompanying increase in $\beta_3$ that smooths the bridge into a quasi-continuous structure between the two main peaks. In Vela and Dragonfly, the higher-altitude pair remains relatively narrow ($\Psi_{c,3}\approx5$--$9^{\circ}$) and Gaussian-like ($\beta_3\approx2$).

Taken together, these trends show that the four-beam geometry reproduces the observed band-to-band variations of the pulse profiles through small to moderate, systematic changes in the plasma-frame beam widths and sharpness, without invoking the extreme energy-dependent beam shapes required in our earlier two-beam  models \citep{Yeung2025}.

\begin{table*}
\small
\centering
\caption{Energy-dependent beam-shape parameters for the Crab, Vela and Dragonfly pulsars.
Best-fitting parameters of the generalised-normal-distribution (GND) beam shapes for Beams 1--2 and Beams 3--4 in each of the four energy bands used in the analysis (60--150~MeV, 150--360~MeV, 360--900~MeV, 0.9--3~GeV). Statistical uncertainties on these parameters are below 1\% and are therefore omitted for clarity.}
\label{tab:beamshape_all}
\begin{tabular}{lcccccccccccc}
\hline
 & \multicolumn{4}{c}{Crab} & \multicolumn{4}{c}{Vela} & \multicolumn{4}{c}{Dragonfly} \\
 & 0.06 & 0.15 & 0.36 & 0.90
 & 0.06 & 0.15 & 0.36 & 0.90
 & 0.06 & 0.15 & 0.36 & 0.90 \\
$E_{\rm det}$ (GeV) & --   & --   & --   & --
 & --   & --   & --   & --
 & --   & --   & --   & --   \\
 & 0.15 & 0.36 & 0.90 & 3.0
 & 0.15 & 0.36 & 0.90 & 3.0
 & 0.15 & 0.36 & 0.90 & 3.0 \\
\hline
\multicolumn{13}{c}{Beams 1 and 2} \\
\hline
$\log_{10} A_{1}$ &
  83.0 & 74.5 & 60.1 & 60.3 &
  28.4 & 36.8 & 43.4 & 56.1 &
  59.2 & 67.6 & 81.2 & 118.6 \\
$\eta_{1}$ &
  340.2 & 299.2 & 237.8 & 238.1 &
   13.2 &  17.5 &  21.2 &  28.1 &
  509.0 & 583.0 & 706.0 & 1049.0 \\
$\Psi_{c,1}$ ($^{\circ}$) &
   2.2 &  2.2 &  2.6 &  2.5 &
  18.6 & 16.3 & 15.0 & 13.4 &
   7.0 &  6.7 &  6.0 &  5.0 \\
$\beta_{1}$ &
  1.38 & 1.35 & 1.34 & 1.33 &
  1.73 & 1.75 & 1.77 & 1.81 &
  1.88 & 1.89 & 1.89 & 1.90 \\
\hline
\multicolumn{13}{c}{Beams 3 and 4} \\
\hline
$\log_{10} A_{3}$ &
  10.9 & 11.0 &  9.5 &  6.4 &
 151.6 &153.4 &145.9 &138.4 &
 128.9 &111.3 & 78.8 & 78.7 \\
$\eta_{3}$ &
  77.4 & 77.8 & 68.5 & 54.0 &
 478.0 &389.0 &369.0 &349.2 &
 532.0 &458.0 &321.2 &321.9 \\
$\Psi_{c,3}$ ($^{\circ}$) &
 10.1 & 10.4 & 13.9 & 30.2 &
  8.7 &  5.4 &  5.6 &  5.8 &
  5.7 &  6.4 &  7.7 &  7.8 \\
$\beta_{3}$ &
 1.33 & 1.35 & 1.45 & 1.94 &
 2.27 & 1.92 & 1.93 & 1.93 &
 1.96 & 1.97 & 1.98 & 2.00 \\
\hline
$C_{\rm bkg}$ &
  893.0 & 704.0 & 424.0 & 218.5 &
  449.0 & 695.0 & 539.0 & 266.6 &
 3100  & 4840  & 3900  & 1998 \\
\hline
\end{tabular}
\end{table*}

\subsection{Beam–motion near alignment}

A striking feature of the fitted solutions is that the bulk-flow directions remain tightly aligned with the corresponding beam axes. For all three pulsars,  the angular separation between the two directions is small to moderate ($\lesssim24^\circ$). This near-alignment suggests that the emitting particles radiate within a narrow pitch-angle cone around the bulk motion, and that any finite cooling length only introduces modest displacements between the geometric beam axis and the direction of maximal Lorentz factor. The geometry therefore supports a picture in which the observed $\gamma$-ray beams trace bulk flows in the 60~MeV--3~GeV range, rather than being strongly smeared by large pitch angles or long propagation distances before emission.

In our framework the fitted beam axis and the fitted bulk-flow direction have related but distinct meanings. The beam-axis angles represent the preferred direction of the $\gamma$-ray emissivity pattern in our corotating geometric coordinate system. The emissivity pattern itself is defined in the local plasma frame, where it is intrinsically brightest along the beam axis. The distinction between the corotating geometric coordinate system and the local plasma frame is clarified in Appendix~\ref{apx:virtual_corot}. The bulk-flow angles instead represent the direction along which individual photons experience the strongest Doppler frequency shift (i.e. the inverse Doppler factor $\varepsilon$ is minimized) when transforming from the local plasma frame to the detector frame. The small fitted misalignments between the two (typically $\lesssim 24^\circ$) are therefore natural, and simply indicate that the emissivity pattern and the kinematic flow are closely aligned but not exactly coincident. Such offsets, preferred by BIC comparisons (Appendix~\ref{apx:crosscheck}), could arise from radiative cooling along the trajectory and from pair production that modulates the emissivity pattern along locally curved magnetic field lines.

\subsection{Speculations on magnetic-field configuration and radiation mechanisms}

Although our framework is explicitly mechanism-independent, the fitted emission geometry and speculated magnetic-field configuration suggests a plausible division of labour between different radiative processes that have been widely discussed in cascade-based pulsar models. In principle, the inferred beam axes approximately trace the local magnetic-field-line directions at the fitted emission sites, because strong cross-field propagation would be more strongly attenuated via magnetic pair creation.

The lower-altitude beam pair, rooted near the last closed field lines and following extended trajectories that graze the light-cylinder boundary, is naturally compatible with curvature-dominated emission from particles moving along large-scale magnetic arcs in the outer magnetosphere (e.g. outer-gap-/slot-gap-type accelerators \citep{Cheng1986, Cheng2000, Hirotani2006, Muslimov2003, Muslimov2004}). The higher-altitude beam pair, by contrast, emerges at larger $|z|$ in regions where the magnetic field is predominantly toroidal (Parker-spiral-like) due to inertial sweepback \citep{Parker1958,Michel1973,Bogovalov1999,Kirk2009} and the flow is more radially outward. Such a configuration  is well suited to synchrotron radiation from relativistic outflow electrons undergoing gyration in a toroidally dominated magnetic field, potentially associated with a striped-wind current-sheet-like structure at high altitudes \citep[e.g.][]{Coroniti1990,Bogovalov1999,Sironi2011,Bai2010,Petri2011,Arka2013,Philippov2015}. 

This division is also reflected in the azimuthal geometry: for Beams~1 and 2 the bulk-flow and beam axes lead their emission sites by almost $90^\circ$ ($\theta_{\rm N},\theta_{\rm Q}\simeq 90^\circ$), indicating nearly tangential motion along closed magnetic arcs, whereas for Beams~3 and 4 the corresponding offsets remain well below $90^\circ$ ($\phi_{\rm N},\phi_{\rm Q}\lesssim 40^\circ$), signalling plasma that has significantly slipped out of azimuthal corotation into  radially outward bulk flow. Within the 60~MeV--3~GeV range considered here, curvature and synchrotron emission in a curved magnetic field can, in principle, account for the lower-altitude and higher-altitude components, respectively. Our results simply identify where in the global flow these two channels are most likely to operate.

\section{\PKHY{Potential extensions of this study}}

The present work focuses on three canonical young pulsars with high-quality Fermi-LAT pulse profiles across 60~MeV--3~GeV, to enable a controlled, like-for-like geometric comparison. The framework can be extended in several directions, each with distinct scientific values:

\begin{itemize}

\item \textbf{Mid-aged and old pulsars.}
Applying the method to mid-aged pulsars (e.g. Geminga, $\sim$300\,kyr) and to recycled millisecond pulsars would test whether the four-beam description persists across different evolutionary and magnetospheric regimes. 

\item \textbf{Full-phase pulsed emission (no off-pulse window).}
For sources where the MeV--GeV pulsed emission may span the full rotation phase \citep{Yeung2025, Abdo_GemingaPSR_2010}, our approach can be adapted to decompose broad pulsed components without prescribing an off-pulse phase \emph{a priori}.

\item \textbf{Multi-wavelength applications.}
Extending the same geometric inference to X-ray and lower-energy phaseograms would enable direct cross-band comparisons of emission-site locations and beam directions, \PKHY{as well as predictions for radio--$\gamma$ phase lags}.

\item \textbf{Polarisation as an external discriminator.}
Phase-dependent polarisation \citep[e.g. in the Crab;][]{Slowikowska2009,Wong_IXPE_2024,Gonzalez2025} can provide independent constraints on the relative contributions and position angles of individual component beams, offering a stringent cross-check on the four-beam decomposition \citep[e.g.][]{Yeung2025}. 

\item \textbf{Inverse-Compton emission at TeV energies.}
For pulsed TeV emission dominated by inverse-Compton scattering (in Cherenkov instruments’ observations), the mapping between seed-photon and up-scattered-photon directions involves successive Lorentz transformations (from the detector frame to the electron rest frame and back). 

\item \textbf{Fainter and single-peak pulsars.}
In lower-S/N or simpler-profile cases, practical progress may require physically motivated priors  and explicit model-comparison criteria to avoid overfitting.

\item \textbf{Magnetic-pole orientation from external modelling.}
An important extension is to connect the fitted emission-site locations and beam directions to the {magnetic pole} geometry predicted by global magnetospheric models. In particular, we may examine the consistency of the inferred four-beam configuration  with a specific magnetic {obliquity} and magnetic-axis {azimuth}.

\item \textbf{Energy-dependent propagation delays.}
Our baseline treatment assumes  negligible propagation/dispersion effects.
Future work could  relax this assumption to test for mild energy-dependent delays from
photon--matter scattering, photon--photon scattering, and/or Lorentz-invariance violation.

\end{itemize}

\section{Summary}

We have presented a mechanism-agnostic, geometry-first four-beam framework for fitting the 60~MeV--3~GeV phaseograms of three canonical young pulsars: Crab, Vela and Dragonfly. Despite their diverse pulse morphologies and energy dependences, the three sources can be organised within a common four-beam template.

In this framework, the phaseograms are described by two beam pairs with distinct geometric and kinematic roles. The lower-altitude pair remains radially close to the light-cylinder boundary and has bulk motion close to azimuthal corotation, naturally accounting for the sharp main peaks. The higher-altitude pair occupies smaller cylindrical radii but higher vertical altitudes, carries a substantial radially outward bulk-motion component, and contributes the bridge/shoulder emission and ripple-like modulations. \PKHY{Such an altitude-dependent azimuthal tilt of pulsed $\gamma$-ray emission could be caused by inertial sweepback in a toroidally dominated (Parker-spiral-like) magnetic field.} For Crab, the phaseogram alone does not uniquely require altitude separation, but co-located solutions are disfavoured when independent constraints on the viewing geometry are imposed (Appendix~\ref{apx:crosscheck}).

Across the three pulsars, the inferred representative height of the higher-altitude pair increases from $\simeq 0.7\,R_{\rm LC}$ in the very young Crab to $\simeq 1.1$--$1.4\,R_{\rm LC}$ in the adolescent Vela and Dragonfly pulsars. \PKHY{Such a contrast may suggest age-dependent vertical inflation of an outflow-transition layer, that needs to be confirmed with a larger sample of broadly aged pulsars}. These qualitative geometric inferences remain broadly robust across the cross-checking analyses and stress tests presented in the Appendix~\ref{apx:crosscheck}.

Although our framework is explicitly mechanism-independent, the inferred geometry is consistent with curvature-dominated outer-magnetospheric emission for the lower-altitude pair and with synchrotron-dominated emission from a toroidally dominated, current-sheet-like outflow for the higher-altitude pair. More broadly, the four-beam framework provides an observationally anchored geometric language for systematic comparison with global MHD/PIC models and for extension to a larger sample of young pulsars.

\begin{acknowledgments}
We are grateful to Dmitry Khangulyan for insightful discussions that helped refine the physical interpretation, including the treatment of phase-dependent Doppler shifts and reference-frame conventions, as well as for advice on the upper energy bound adopted for the global analysis and for recommendations on wording.
Sincere gratitude is given to Matthew Kerr for providing an optimised timing ephemeris for the Vela pulsar. We are grateful to the Fermi-LAT collaboration for operating the mission and providing public access to the data. We acknowledge the H.E.S.S. Collaboration’s study of the Vela pulsar, which provides valuable context for interpreting the wind-regime geometry discussed here \citep{HESS_2023_VelaPSR}. We also acknowledge the pioneering outer-magnetospheric gap models proposed by  \cite{Arons1979} and  \cite{Cheng1986}. In addition, we note the foundational discussion of emission at radii extending beyond the light cylinder by  \cite{Aharonian2012}. Thanks are also given to the anonymous referee for the fruitful comments. Last but not least, we thank Ievgen Vovk for useful discussion.
\end{acknowledgments}

\begin{contribution}

P. K. H. Yeung developed the fitting framework and software, performed the phaseogram fits and analysis, and drafted the manuscript. 
T. Saito supervised the project and advised on cross-check analyses. All authors reviewed and edited the manuscript.


\end{contribution}

%
\facilities{Fermi-LAT}

\software{iminuit}


\appendix
\twocolumngrid
\setcounter{table}{0}
\renewcommand{\thetable}{S\arabic{table}}
\setcounter{figure}{0}
\renewcommand{\thefigure}{S\arabic{figure}}
\setcounter{equation}{0}
\renewcommand{\theequation}{S\arabic{equation}}

\section{Azimuthal separation between the lower-altitude and higher-altitude emission sites}\label{apx:az_sep}

The azimuthal location of each emission site around the spin axis is encoded in the
phase parameters $t_0$ (for Beams~1 and 2) and $T_0$ (for Beams~3 and 4), which
denote the detector-frame phases at which the corresponding emission site passes
closest to our line of sight.  Using the light-travel-time transformation between
the pulsar and detector frames, as given by Equations~(2) and (4) of \cite{Yeung2025}, the azimuthal offset between the lower-altitude and higher-altitude emission sites can
be expressed in terms of the fitted quantities
$\Theta_{\mathrm{A}}$, $d\sin\theta_{\mathrm{B}}$, $D\sin\phi_{\mathrm{B}}$, $t_0$ and $T_0$:
\begin{equation}\label{DeltaAzSite}
\begin{split}
\Delta\varphi_{\rm site}
 &= 360^{\circ}\left[
   (T_0 - t_0)\right.\\
 &\quad \left.
   + \frac{d\cos(\Theta_{\rm A}+\theta_{\rm B})
          - D\cos(\Theta_{\rm A}+\phi_{\rm B})}{2\pi\RLC}
   \right].
\end{split}
\end{equation}
For each pulsar we evaluate this expression a posteriori, so that the radial and
vertical separations quoted in the main text are complemented by the full
three-dimensional $({\Delta}r, {\Delta}z, {\Delta}\varphi_{\rm site})$ offset between the two emission
regions. Geometrically, a value of $\Delta\varphi_{\rm site}$ in the range
$0^\circ$ to $180^\circ$ means that the higher-altitude emission site reaches its
closest approach to the line of sight at a later pulsar-frame time than the
lower-altitude site, whereas a value in the range $-180^\circ$ to $0^\circ$
indicates that it does so earlier. In addition, the azimuthal separation between the corresponding beam
axes is obtained by combining Eq.~\ref{DeltaAzSite} with the fitted azimuthal offsets
$\theta_{\mathrm{Q}}$ and $\phi_{\mathrm{Q}}$, yielding:
\begin{equation}\label{DeltaAzBeam}
\Delta\varphi_{\rm beam}
= \Delta\varphi_{\rm site} - (\phi_{\rm Q} - \theta_{\rm Q}) .
\end{equation}

\section{\PKHY{Further interpretation of $\eta_i$}}\label{apx:eta_i}

Because of the  many-to-many (i.e. neither injective nor surjective) correspondence between $E_{\rm bulk}$ and $E_{\rm det}$ \citep[Section~5 of][]{Yeung2025}, a fixed detector-energy band samples a \emph{distribution} of plasma-frame energies whose support and relative weighting can change with rotational phase through the inverse Doppler factor $\varepsilon(\varphi)$.
As a result, beam-shape parameters inferred from phaseograms in separate $E_{\rm det}$ bands are not required to evolve smoothly or monotonically with energy: modest band-to-band changes can naturally arise from shifts in the effective $E_{\rm bulk}$ weighting, even if the underlying comoving emissivity varies smoothly.

\PKHY{According to \S\ref{main:eta_i},} although $\eta_i$ is not intended to be interpreted as a photon index, its construction is \emph{interface-friendly} to physical models in the following sense: any specific emission calculation (outer-gap, current-sheet, PIC-informed prescriptions, etc.) predicts a comoving emissivity $j(E_{\rm bulk},\psi)$, which can be forward-folded through the same Doppler mapping and finite $E_{\rm det}$ bandpass used here to generate synthetic phaseograms. Fitting those synthetic phaseograms with our parametric form yields the corresponding effective parameters $(\Psi_{c,i},\beta_i,\eta_i)$ for like-for-like comparison with the data.
In simple limiting cases where the comoving spectrum over the relevant band is close to a power law and the angular kernel is only weakly energy dependent, the band-limited intensity acquires an approximate power-law dependence on $\varepsilon$ whose exponent is set by the local spectral slope (up to convention-dependent constants), so $\eta_i$ can be viewed as a compact \emph{summary statistic} of how the model's band-integrated emissivity responds to Doppler reweighting, including any residual energy dependence of the angular pattern.

Furthermore, because our fits use band-limited photon counts, the positive sign of $\eta_i$ does not encode a universal forward beaming of photon number. Instead, it captures the band-integrated Doppler response, which depends on the local spectral slope/curvature. For spectra with strong curvature or cutoffs, the band-limited photon counts can be preferentially enhanced at directions deviating from the bulk-motion direction (i.e. at viewing directions corresponding to larger $\varepsilon$).

\section{Segregation between pattern-level ``virtual" corotation and kinematic bulk flow}\label{apx:virtual_corot}

In our previous study (Appendix~C of \cite{Yeung2025}) we introduced the notion of a “virtual motion’’ of the emission site. The key idea is that, at a given geometric location in a corotating frame, the photon-emitting particles are constantly replaced in a wave-like manner. The population of particles constituting the emission region at one spin phase is therefore different from (though spatially overlapping with) the population at another spin phase. What appears observationally as a “corotating’’ emission site is in fact a pattern of emissivity sweeping across space, rather than a single cloud of particles rigidly corotating with the star. Because this pattern is carried by continuous replacement of particles, the apparent speed of such a virtual motion is not limited by the speed of light and can remain phase-locked to the pulsar’s rotation, even when the representative emission radius exceeds the light cylinder.

Here we explicitly segregate the virtual motion of the emissivity pattern from the kinematic bulk flow of the emitting plasma, and likewise the corotating frame from the plasma frame. The bulk flow sets the inverse Doppler factor $\varepsilon$, which relates detector-frame photon energies to those in the plasma frame, i.e. the frame where the average drift of the emission region vanishes. The periodic modulation seen in the detector frame, as well as the effective light-travel-time delays, is instead governed by the virtual corotation of the emissivity pattern.

In the present modelling framework we generalise this construction to emission sites with fitted cylindrical radii both below and above $R_{\rm LC}$. In addition to a Lorentz factor $\gamma_{\rm Lor}$, we parametrise the bulk flow for each beam by a set of orientation angles that determine the angle $\psi_{\rm vel}$ between the bulk-velocity vector and the line of sight. In this way, even emission sites with $r<R_{\rm LC}$ are not assumed to be in rigid corotation. Without requiring rigid corotation of the plasma or any superluminal material motion, our model keeps the roles of virtual motion (pattern-level synchronisation, pulsation and light-travel-time delays) and kinematic bulk flow (Doppler environment) strictly segregated. Related constructions and specific examples can be found in previous work \citep[e.g.][]{Aharonian2012,HESS_2023_VelaPSR}.

In the context of pulsars, even a so-called plasma frame is strictly local and momentary, because the continuous replacement of photon-emitting particles is accompanied by a continuous replacement of the comoving frame. In principle, each parcel of plasma radiates only for a short time and then ceases to contribute, rather than forming a long-lived, rigidly emitting cloud. Nevertheless, we stress that Equation~(\ref{eqn:Doppler}) in this work is defined with respect to this ever-replacing local plasma frame, so that the Doppler factor always refers to the instantaneous comoving plasma that is actually emitting at a given phase.

\section{Crosschecking analyses, BIC comparisons \& stress tests}\label{apx:crosscheck}

\subsection{BIC model comparison: two-beam versus four-beam decomposition}

We compare the two-beam and four-beam models using the Bayesian information criterion (BIC),
${\rm BIC}=k\ln n-2\ln\mathcal{L}_{\max}$, where $k$ is the number of free parameters ($k=29$ for the two-beam model; $k=53$ for the four-beam model) and
$n$ is the total number of phase bins across all energy bands used in the joint fit.
The results are summarised in Table~\ref{tab:BIC_stat}. Using the tabulated quantities, $\Delta{\rm BIC}=2\ln(\mathcal{L}_{\max,4{\rm b}}/\mathcal{L}_{\max,2{\rm b}})-24\ln n$.
In all three pulsars, the four-beam model is strongly preferred ($\Delta{\rm BIC}>600$).
Given this decisive preference and the increasing degeneracy expected when introducing additional components,
we do not explore higher-order decompositions.

\begin{table}[t]
\small
\setlength{\tabcolsep}{2pt}
\centering
\caption{BIC statistics for comparing the two-beam and four-beam models.
Here $n$ is the total number of phase bins across all fitted phaseograms,
$k$ is the number of free parameters, and $\ln\mathcal{L}_{\max}$ is the maximum joint Poisson log-likelihood.
We define $\Delta{\rm BIC}\equiv {\rm BIC}_{2{\rm b}}-{\rm BIC}_{4{\rm b}}$, so that $\Delta{\rm BIC}>0$
favours the four-beam model.}
\label{tab:BIC_stat}
\begin{tabular}{@{}lccc@{}}
\hline
     & Crab & Vela & Dragonfly \\
\hline
$2\ln(\mathcal{L}_{\max,4{\rm b}}/\mathcal{L}_{\max,2{\rm b}})$ &  $3.7\times10^{3}$ & $3.8\times10^{4}$ & $7.6\times10^{2}$  \\
$n$ & 1000 & 2000 & 320 \\
$k_{4{\rm b}}-k_{2{\rm b}}$ & 24 & 24 & 24 \\
$\Delta{\rm BIC}$ &  $3.5\times10^{3}$ & $3.8\times10^{4}$ & $6.3\times10^{2}$  \\
\hline
\end{tabular}
\end{table}

\subsection{BIC model comparison: escaping plasma versus azimuthal-motion limit}

We test whether the bulk-motion directions inferred in our baseline (escaping-plasma) scenario
are required by the phaseograms, or whether they are consistent with the azimuthal-motion
(rotation) limit, in which the bulk flow is purely tangential.
In this limit we fix the bulk-motion angles to
$\theta_M=\theta_N=\phi_M=\phi_N=90^\circ$,
and refit the data with all remaining parameters left free.

We compare the two variants using the BIC.
The results are summarised in Table~\ref{tab:BIC_esc_az}.
We define $\Delta{\rm BIC}\equiv {\rm BIC}_{\rm az}-{\rm BIC}_{\rm esc}$, so that
$\Delta{\rm BIC}>0$ favours the escaping-plasma model.
Using the tabulated quantities,
$\Delta{\rm BIC}
=2\ln\!\left(\mathcal{L}_{\max,{\rm esc}}/\mathcal{L}_{\max,{\rm az}}\right)
-4\ln n$.
In all three pulsars, the escaping-plasma model is strongly preferred ($\Delta{\rm BIC}>200$).

We further test a hybrid variant in which one beam pair is restricted to the azimuthal-motion (purely tangential) limit by fixing its bulk-flow angles to 
$\theta_M=\theta_N=90^\circ$ (with the other pair’s bulk-flow angles left free), and compare it to the baseline using the BIC.
Using the tabulated quantities,
$\Delta{\rm BIC}
=2\ln\!\left(\mathcal{L}_{\max,{\rm esc}}/\mathcal{L}_{\max,{\rm hyb}}\right)
-2\ln n$.
As a secondary plausibility check, we also report the fitted $\Theta_A$ values for the escaping-plasma model and the hybrid variant, and the
independently inferred $\zeta$ from PWN torus modelling (Table~\ref{tab:BIC_esc_az}). While not used in the BIC,
$\Theta_A$ is expected to be broadly consistent with $\zeta$ within modelling uncertainties.

The BIC preference of the escaping-plasma model over the hybrid variant is overwhelming for Vela
($\Delta{\rm BIC}=1.9\times10^{4}$),
and is also significant for Crab and Dragonfly
($\Delta{\rm BIC}=19$--35).
Additionally, within the hybrid scenario, the statistically best-fitting solutions for Crab and Dragonfly
require $\Theta_{\rm A}$ to deviate substantially from $\zeta$.
Taken together, the BIC tests and the $\Theta_{\rm A}$--$\zeta$ comparison indicate that the pulse profiles
are best explained when neither pulse-emitting population is restricted to purely tangential bulk motion,
favouring escaping-flow geometries in all three pulsars.

\begin{table}[t]
\small
\setlength{\tabcolsep}{2pt}
\centering
\caption{BIC statistics for comparing the escaping-plasma model, the azimuthal-motion
(rotation) limit and the hybrid variant.
Here $n$ is the total number of phase bins across all fitted phaseograms,
$k$ is the number of free parameters, and $\ln\mathcal{L}_{\max}$ is the maximum
joint Poisson log-likelihood.
We define $\Delta{\rm BIC}\equiv {\rm BIC}_{\rm az/hyb}-{\rm BIC}_{\rm esc}$, so that
$\Delta{\rm BIC}>0$ favours the escaping-plasma model.
We also report $\Theta_A$ for the escaping-plasma model and the hybrid variant, and the independently inferred $\zeta$ from PWN torus modelling.}
\label{tab:BIC_esc_az}
\begin{tabular}{@{}lccc@{}}
\hline
 & Crab & Vela & Dragonfly \\
\hline
$2\ln(\mathcal{L}_{\max,{\rm esc}}/\mathcal{L}_{\max,{\rm az}})$ &  $3.1\times10^{2}$ & $3.5\times10^{4}$ & $2.3\times10^{2}$  \\
$2\ln(\mathcal{L}_{\max,{\rm esc}}/\mathcal{L}_{\max,{\rm hyb}})$ &  $4.9\times10^{1}$ & $1.9\times10^{4}$ & $3.1\times10^{1}$  \\
$n$ & 1000 & 2000 & 320 \\
$k_{\rm esc}-k_{\rm az}$ & 4 & 4 & 4 \\
$k_{\rm esc}-k_{\rm hyb}$ & 2 & 2 & 2 \\
$\Delta{\rm BIC}_{\rm az}$ &  $2.9\times10^{2}$ & $3.5\times10^{4}$ & $2.1\times10^{2}$  \\
$\Delta{\rm BIC}_{\rm hyb}$ &  $3.5\times10^{1}$ & $1.9\times10^{4}$ & $1.9\times10^{1}$  \\
$\Theta_{A,{\rm esc}}\ (^\circ)$ & 55.7 & 54.9 & 80.9 \\
$\Theta_{A,{\rm hyb}}\ (^\circ)$ & 33.1 & 52.8 & 68.0 \\
$\zeta^\ast\ (^\circ)$ & 61--63 & 63.6 & 85--88 \\
\hline
\end{tabular}

\footnotesize{\textit{Notes.} $^\ast$ The inclination angle between the polar axis of the PWN’s equatorial torus and our line of sight
\citep[e.g.][]{Ng2004,VanEtten2008,Jin2023}.}
\end{table}

\subsection{BIC model comparison: beam--motion total alignment versus slight misalignment}

Our baseline framework allows a misalignment between the fitted beam axis and the fitted bulk-flow direction. The beam axis is the preferred direction of the emissivity pattern, parameterised in our corotating geometric coordinate system. Whereas, the bulk-flow direction corresponds to the strongest Doppler frequency shift for individual photon energies (i.e. $\varepsilon$ is minimized) when transforming from the local plasma frame to the detector frame.
To test whether our obtained slight misalignment ($\lesssim24^\circ$) is required by the phaseograms, we construct a restricted
``total-alignment'' variant in which the bulk-flow direction is forced to coincide with the beam axis
for each component beam.

Specifically, in the total-alignment limit we fix the bulk-motion angles to the corresponding
beam-axis angles
(i.e. $\theta_M=\theta_C$, $\theta_N=\theta_Q$, $\phi_M=\phi_C$, and $\phi_N=\phi_Q$),
and refit the data with all remaining parameters left free.
This constraint removes four degrees of freedom relative to the baseline model.

We compare the baseline and total-alignment variants using the BIC.
The results are summarised in Table~\ref{tab:BIC_mis_align}.
We define $\Delta{\rm BIC}\equiv {\rm BIC}_{\rm align}-{\rm BIC}_{\rm base}$, so that
$\Delta{\rm BIC}>0$ favours the baseline model with slight beam--motion misalignment.
Using the tabulated quantities,
$\Delta{\rm BIC}
=2\ln\!\left(\mathcal{L}_{\max,{\rm base}}/\mathcal{L}_{\max,{\rm align}}\right)
-4\ln n$,
where $n$ is the total number of fitted phase bins across all phaseograms.
In all three pulsars, the baseline model is overwhelmingly preferred, with
$\Delta{\rm BIC}>350$, demonstrating that the phaseograms strongly favour a small but non-zero
beam--motion misalignment within our geometric description.

\begin{table}[t]
\small
\setlength{\tabcolsep}{2pt}
\centering
\caption{BIC statistics for comparing the baseline model (allowing slight beam--motion misalignment)
and the total-alignment limit.
Here $n$ is the total number of phase bins across all fitted phaseograms,
$k$ is the number of free parameters, and $\ln\mathcal{L}_{\max}$ is the maximum
joint Poisson log-likelihood.
We define $\Delta{\rm BIC}\equiv {\rm BIC}_{\rm align}-{\rm BIC}_{\rm base}$, so that
$\Delta{\rm BIC}>0$ favours the baseline model.}
\label{tab:BIC_mis_align}
\begin{tabular}{@{}lccc@{}}
\hline
 & Crab & Vela & Dragonfly \\
\hline
$2\ln(\mathcal{L}_{\max,{\rm base}}/\mathcal{L}_{\max,{\rm align}})$ &  $1.2\times10^{3}$ & $2.0\times10^{4}$ & $3.8\times10^{2}$  \\
$n$ & 1000 & 2000 & 320 \\
$k_{\rm base}-k_{\rm align}$ & 4 & 4 & 4 \\
$\Delta{\rm BIC}$ &   $1.2\times10^{3}$ & $2.0\times10^{4}$ & $3.5\times10^{2}$  \\
\hline
\end{tabular}
\end{table}

\subsection{BIC model comparison: energy-dependent versus energy-independent sharpness indices}

In our baseline implementation, the GND sharpness indices $\beta_1$ and $\beta_3$ are allowed to vary between the four $E_{\rm det}$ bands (Table~\ref{tab:beamshape_all}), providing flexibility for modest energy-dependent changes in the beam-edge sharpness.
To test whether such band-to-band variations are required by the phaseograms, we construct a restricted
``energy-independent-$\beta$'' variant in which $\beta_1$ and $\beta_3$ are each forced to be globally uniform across all four $E_{\rm det}$ bands for a given pulsar, while all remaining parameters are refitted.
This constraint reduces the number of free parameters by six relative to the baseline model (replacing 8 band-dependent sharpness parameters by 2 band-independent constants).

We compare the baseline and energy-independent-$\beta$ variants using the BIC.
The results are summarised in Table~\ref{tab:BIC_beta_const}.
We define $\Delta{\rm BIC}\equiv {\rm BIC}_{\beta\text{-}{\rm const}}-{\rm BIC}_{\rm base}$, so that
$\Delta{\rm BIC}>0$ favours the baseline model with energy-dependent sharpness indices.
Using the tabulated quantities,
$\Delta{\rm BIC}
=2\ln\!\left(\mathcal{L}_{\max,{\rm base}}/\mathcal{L}_{\max,\beta\text{-}{\rm const}}\right)
-6\ln n$,
where $n$ is the total number of fitted phase bins across all phaseograms.
As a secondary plausibility check, we also report the fitted $\Theta_A$ values for both variants, and the
independently inferred $\zeta$ from PWN torus modelling (Table~\ref{tab:BIC_beta_const}). While not used in the BIC,
$\Theta_A$ is expected to be broadly consistent with $\zeta$ within modelling uncertainties.

The BIC preference is strong for Crab and Vela ($\Delta{\rm BIC}>300$) and modest for Dragonfly ($\Delta{\rm BIC}=18$).
The $\Theta_A$ values provide additional support for the baseline choice: in all three sources, enforcing
energy-independent $\beta$ shifts $\Theta_A$ away from the independently inferred $\zeta$, most notably in Crab and
Dragonfly. In particular, the energy-independent-$\beta$ variant yields $\Theta_A=43.3^\circ$ (Crab) and $71.5^\circ$
(Dragonfly), substantially below $\zeta=61$--$63^\circ$ and $85$--$88^\circ$, respectively, whereas the baseline
solutions are closer to these PWN-based estimates.
Taken together, for all three pulsars, the BIC comparison and the $\Theta_A$ plausibility check favour retaining modest band-to-band flexibility in the sharpness indices within our framework.

\begin{table}[t]
\small
\setlength{\tabcolsep}{2pt}
\centering
\caption{BIC statistics for comparing the baseline model (allowing band-dependent sharpness indices $\beta_1$ and $\beta_3$) and an energy-independent-$\beta$ variant in which $\beta_1$ and $\beta_3$ are each globally uniform across all four $E_{\rm det}$ bands for a given pulsar, with all other parameters refitted.
Here $n$ is the total number of phase bins across all fitted phaseograms,
$k$ is the number of free parameters, and $\ln\mathcal{L}_{\max}$ is the maximum
joint Poisson log-likelihood.
We define $\Delta{\rm BIC}\equiv {\rm BIC}_{\beta\text{-}{\rm const}}-{\rm BIC}_{\rm base}$, so that
$\Delta{\rm BIC}>0$ favours the baseline model.
We also report $\Theta_A$ for both
variants, and the independently inferred $\zeta$ from PWN torus modelling.}
\label{tab:BIC_beta_const}
\begin{tabular}{@{}lccc@{}}
\hline
 & Crab & Vela & Dragonfly \\
\hline
$2\ln(\mathcal{L}_{\max,{\rm base}}/\mathcal{L}_{\max,\beta\text{-}{\rm const}})$ &  371 & 3640 & 53 \\
$n$ & 1000 & 2000 & 320 \\
$k_{\rm base}-k_{\beta\text{-}{\rm const}}$ & 6 & 6 & 6 \\
$\Delta{\rm BIC}$ & 329 & 3594 & 18 \\
$\Theta_{A,{\rm base}}\ (^\circ)$ & 55.7 & 54.9 & 80.9 \\
$\Theta_{A,\beta\text{-}{\rm const}}\ (^\circ)$ & 43.3 & 49.9 & 71.5 \\
$\zeta^\ast\ (^\circ)$ & 61--63 & 63.6 & 85--88 \\
\hline
\end{tabular}

\footnotesize{\textit{Notes.} $^\ast$ The inclination angle between the polar axis of the PWN’s equatorial torus and our line of sight
\citep[e.g.][]{Ng2004,VanEtten2008,Jin2023}.}
\end{table}

\subsection{BIC model comparison: altitude-separated beam pairs versus the co-location limit}

In our baseline four-beam geometry, the lower-altitude and higher-altitude beam pairs are allowed to have
independent location parameters (and hence can be altitude-separated).
To test whether such a separation is required by the phaseograms, we construct a restricted
``co-location'' variant in which the two beam pairs are forced to share the same location parameters by fixing
$D=d$, $\phi_B=\theta_B$, and $T_0=t_0$, and refit the data with all remaining parameters left free.
This constraint removes three degrees of freedom relative to the baseline model.

We compare the baseline and co-location variants using the Bayesian information criterion (BIC).
The results are summarised in Table~\ref{tab:BIC_coloc}.
We define $\Delta{\rm BIC}\equiv {\rm BIC}_{\rm co\text{-}loc}-{\rm BIC}_{\rm base}$, so that
$\Delta{\rm BIC}>0$ favours the baseline model.
Using the tabulated quantities,
$\Delta{\rm BIC}
=2\ln\!\left(\mathcal{L}_{\max,{\rm base}}/\mathcal{L}_{\max,{\rm co\text{-}loc}}\right)
-3\ln n$,
where $n$ is the total number of fitted phase bins across all phaseograms.

The baseline model is strongly preferred for Vela ($\Delta{\rm BIC}=312.8$). For Dragonfly, the two variants are
statistically indistinguishable ($\Delta{\rm BIC}=0.5$).
Figure~\ref{fig:phaseograms} shows that Beam~4 contributes only weakly in all bands, which limits the constraints
on the higher-altitude pair and may contribute to the small $\Delta{\rm BIC}$.
For Crab, the BIC shows a weak preference for the co-location limit ($\Delta{\rm BIC}=-7.9$), indicating that the
phaseograms alone do not decisively require altitude separation in this source.

As a secondary plausibility check, we compare the fitted $\Theta_A$ values (the fitted angle between our line of sight and the pulsar's rotational axis) with the independently inferred
$\zeta$ from PWN torus modelling (Table~\ref{tab:BIC_coloc}).
For Crab, within the co-location limit, statistically best-fitting solutions require $\Theta_A$ to deviate substantially from $\zeta$. Imposing $\Theta_A\sim\zeta$ increases the BIC by $\gtrsim$97.9 relative to the baseline model. For Vela, $\zeta$ also favours the baseline
model over the co-location limit.

Taken together, these two diagnostics, based on BIC and $\Theta_A$ respectively, indicate that relaxing the
co-location constraints is warranted to achieve a self-consistent geometric description across the three pulsars.

\begin{table}[t]
\small
\setlength{\tabcolsep}{2pt}
\centering
\caption{BIC statistics for comparing the baseline model and the co-location limit in which $D=d$, $\phi_B=\theta_B$,
and $T_0=t_0$ are fixed and all other parameters are refitted.
Here $n$ is the total number of phase bins across all fitted phaseograms, $k$ is the number of free parameters, and
$\ln\mathcal{L}_{\max}$ is the maximum joint Poisson log-likelihood.
We define $\Delta{\rm BIC}\equiv {\rm BIC}_{\rm co\text{-}loc}-{\rm BIC}_{\rm base}$, so that $\Delta{\rm BIC}>0$
favours the baseline model.
We also report $\Theta_A$ for both
variants, and the independently inferred $\zeta$ from PWN torus modelling.}
\label{tab:BIC_coloc}
\begin{tabular}{@{}lcccc@{}}
\hline
 & \multicolumn{2}{c}{Crab} & Vela & Dragonfly \\
\hline
$2\ln(\mathcal{L}_{\max,{\rm base}}/\mathcal{L}_{\max,{\rm co\text{-}loc}})$ & 12.8 & \{125.6\} & 335.7 & 17.8 \\
$n$ & \multicolumn{2}{c}{1000} & 2000 & 320 \\
$k_{\rm base}-k_{\rm co\text{-}loc}$ & 3 & \{4\} & 3 & 3 \\
$\Delta{\rm BIC}$ & $-7.9$ & \{97.9\} & 312.8 & 0.5 \\
$\Theta_{A,{\rm base}}\ (^\circ)$ & \multicolumn{2}{c}{55.7} & 54.9 & 80.9 \\
$\Theta_{A,{\rm co\text{-}loc}}\ (^\circ)$ & 21.0 & \{50\}$^\dag$ & 46.4 & 83.6 \\
$\zeta^\ast\ (^\circ)$ & \multicolumn{2}{c}{61--63} & 63.6 & 85--88 \\
\hline
\end{tabular}

\footnotesize{\textit{Notes.} $^\ast$ The inclination angle between the polar axis of the PWN’s equatorial torus and our line of sight
\citep[e.g.][]{Ng2004,VanEtten2008,Jin2023}.\\
$^\dag$ For Crab, values in curly brackets correspond to a constrained co-location fit with $\Theta_A$ fixed at $50^\circ$ (chosen close to $\zeta$), while values outside brackets correspond to the co-location fit with $\Theta_A$ free. }
\end{table}

\subsection{Constraints on the bulk Lorentz factors and sensitivity saturation at high $\gamma_{\rm Lor}$}

Our baseline fits (Table~\ref{tab:geometry}) infer bulk Lorentz factors $\gamma_{\rm Lor,1}$ and $\gamma_{\rm Lor,3}$
for the lower- and higher-altitude beam pairs, respectively.
To test whether these fitted values might be underestimated, we construct restricted higher-Lorentz-factor variants
in which both Lorentz factors are fixed to $f$ times their baseline best-fitting values (with $f=1.2,1.5,2,2.5,3,9,90,900$),
and refit the data with all remaining parameters left free. Each test reduces the number of free parameters by two
relative to the baseline model.

We compare the baseline and higher-Lorentz-factor variants using the BIC.
The results are summarised in Table~\ref{tab:BIC_highgamma}.
We define $\Delta{\rm BIC}\equiv {\rm BIC}_{f\times{\rm Lor}}-{\rm BIC}_{\rm base}$, so that $\Delta{\rm BIC}>0$
favours the baseline model. Using the tabulated quantities,
$\Delta{\rm BIC}=2\ln\!\left(\mathcal{L}_{\max,{\rm base}}/\mathcal{L}_{\max,f\times{\rm Lor}}\right)-2\ln n$,
where $n$ is the total number of fitted phase bins across all phaseograms.

The resulting $\Delta{\rm BIC}$ values are not strictly monotonic in $f$, indicating parameter degeneracy (and hence multiple nearby solutions) within the restricted high-$\gamma_{\rm Lor}$ variants.
Despite this degeneracy, every restricted variant still yields $\Delta{\rm BIC}\ge16$, consistently
favouring the baseline Lorentz factors reported in Table~\ref{tab:geometry}. We therefore conclude that the fitted
Lorentz factors do not appear to be underestimated within our geometric description.

For Crab and Dragonfly, this comparison suggests that plasma with highly relativistic bulk motion is not favoured as
the origin of the observed pulsed $\gamma$-ray emission within our framework. In particular, a conservative upper
limit on the bulk Lorentz factor of the pulse-emitting plasma is $\gamma_{\rm Lor}\lesssim1.5$ for these two pulsars,
favouring mildly relativistic values. For Dragonfly the preference is modest ($\Delta{\rm BIC}=16$--23), so we treat
this upper limit as conservative rather than definitive. This does not preclude the presence of more highly relativistic
plasma in these systems. It constrains only the component producing the pulsed GeV emission captured by our model.

At large multipliers ($f\gtrsim9$), the $\Delta{\rm BIC}$ values approach a plateau rather than continuing to increase, indicating that the model sensitivity to further Lorentz-factor enhancement has effectively saturated and that the parameter degeneracy is accompanied by convergence at $\gamma_{\rm Lor}\gtrsim10$.
This behaviour is also expected analytically from Equation~(\ref{eqn:Doppler}), where the inverse Doppler factor
$\varepsilon \equiv E_{\rm bulk}/E_{\rm det}$ depends on the bulk motion only through the combination
$\gamma_{\rm Lor}(1-\frac{v}{c}\cos\psi_{\rm vel})$ in the denominator.
In the highly relativistic regime, $v\rightarrow{c}$ and variations in $\gamma_{\rm Lor}$ mainly rescale
$\varepsilon$ approximately as $\varepsilon\propto 1/\gamma_{\rm Lor}$ for fixed $\psi_{\rm vel}$ (except for near head-on geometries with $\psi_{\rm vel}\approx0$).
Consequently, once $\gamma_{\rm Lor}$ is pushed to sufficiently large values, further increases primarily produce an overall rescaling of $\varepsilon$ (approximately $\propto 1/\gamma_{\rm Lor}$) that can be partially absorbed by refitting the emissivity normalisations and the power-law response to $\varepsilon$ (through $\eta$), rather than generating qualitatively new phase-dependent structure.
This naturally leads to diminishing returns in the likelihood improvement and therefore a convergent (plateauing) $\Delta{\rm BIC}$ trend at large $f$, consistent with the numerical behaviour seen for $f=9$, $90$ and $900$ in Table~\ref{tab:BIC_highgamma}.

This plateauing behaviour also highlights a practical limitation of our framework: once the bulk motion enters the highly relativistic regime, the phaseograms cease to provide strong leverage on the absolute value of $\gamma_{\rm Lor}$ beyond $\sim$ a few--ten. 
In this regime, the model sensitivity is dominated by the combination
$\gamma_{\rm Lor}\!\left(1-\frac{v}{c}\cos\psi_{\rm vel}\right)$ in Equation~(\ref{eqn:Doppler}).
For $\gamma_{\rm Lor}\gtrsim10$ (i.e. $v/c\gtrsim0.995$) and for angles not extremely close to
$\psi_{\rm vel}=0$, this dependence is well approximated by $\gamma_{\rm Lor}(1-\cos\psi_{\rm vel})$.
Consequently, very large best-fitting values should be interpreted cautiously: for example, the formal $\gamma_{\rm Lor,1}\simeq37.3$ obtained for Vela likely reflects that, once $\gamma_{\rm Lor,1}\gtrsim10$, the likelihood becomes sufficiently flat that small unmodelled effects (e.g. instrumental response residuals or timing-noise-related imperfections in the phaseograms) can shift the maximum-likelihood point without implying a well-measured ultra-high Lorentz factor. We therefore interpret this result primarily as evidence for strongly relativistic bulk motion in the lower-altitude component of Vela, corresponding to a conservative lower limit $\gamma_{\rm Lor,1}\gtrsim10$, while the data do not provide a meaningful upper bound within our current geometric description. Importantly, this limitation concerns only the absolute scale of $\gamma_{\rm Lor}$ in the ultra-relativistic regime and does not affect the qualitative geometric conclusions drawn from the emission-site locations and their altitude stratification.

\begin{table}[t]
\small
\setlength{\tabcolsep}{2pt}
\centering
\caption{BIC statistics for exploring higher-Lorentz-factor variants in which the Lorentz factors
$\gamma_{\rm Lor,1}$ and $\gamma_{\rm Lor,3}$ are fixed to $f$ times their baseline best-fitting values
(Table~\ref{tab:geometry}), with $f=1.2,1.5,2,2.5,3,9,90,900$ and all other parameters refitted.
Here $n$ is the total number of phase bins across all fitted phaseograms, $k$ is the number of free parameters, and $\ln\mathcal{L}_{\max}$ is the maximum
joint Poisson log-likelihood. We define $\Delta{\rm BIC}\equiv {\rm BIC}_{f\times{\rm Lor}}-{\rm BIC}_{\rm base}$,
so that $\Delta{\rm BIC}>0$ favours the baseline model.}
\label{tab:BIC_highgamma}
\begin{tabular}{@{}lccc@{}}
\hline
 & Crab & Vela & Dragonfly \\
\hline
$2\ln(\mathcal{L}_{\max,{\rm base}}/\mathcal{L}_{\max,{\rm 1.2{\times}Lor}})$ &   130 & 1721 & 35   \\
$2\ln(\mathcal{L}_{\max,{\rm base}}/\mathcal{L}_{\max,{\rm 1.5{\times}Lor}})$ &   95 & 4078 & 32   \\
$2\ln(\mathcal{L}_{\max,{\rm base}}/\mathcal{L}_{\max,{\rm 2{\times}Lor}})$ &  96 & 4416 & 28   \\
$2\ln(\mathcal{L}_{\max,{\rm base}}/\mathcal{L}_{\max,{\rm 2.5{\times}Lor}})$ &  89 & 4487 & 27   \\
$2\ln(\mathcal{L}_{\max,{\rm base}}/\mathcal{L}_{\max,{\rm 3{\times}Lor}})$ &  88 & 5556 & 28   \\
$2\ln(\mathcal{L}_{\max,{\rm base}}/\mathcal{L}_{\max,{\rm 9{\times}Lor}})$ &  85 & 5715 & 28   \\
$2\ln(\mathcal{L}_{\max,{\rm base}}/\mathcal{L}_{\max,{\rm 90{\times}Lor}})$ &  84 & 5735 & 27   \\
$2\ln(\mathcal{L}_{\max,{\rm base}}/\mathcal{L}_{\max,{\rm 900{\times}Lor}})$ &  84 & 5735 & 27   \\
$n$ & 1000 & 2000 & 320 \\
$k_{\rm base}-k_{f\times{\rm Lor}}$ & 2 & 2 & 2 \\
$\Delta{\rm BIC}_{\rm 1.2{\times}Lor}$ &  117 & 1706 & 23 \\
$\Delta{\rm BIC}_{\rm 1.5{\times}Lor}$ &  81 & 4063 & 21 \\
$\Delta{\rm BIC}_{\rm 2{\times}Lor}$ & 82 & 4401 & 16 \\
$\Delta{\rm BIC}_{\rm 2.5{\times}Lor}$ & 76 & 4472 & 16 \\
$\Delta{\rm BIC}_{\rm 3{\times}Lor}$ & 74 & 5540 & 16 \\
$\Delta{\rm BIC}_{\rm 9{\times}Lor}$ & 71 & 5699 & 16 \\
$\Delta{\rm BIC}_{\rm 90{\times}Lor}$ & 71 & 5720 & 16 \\
$\Delta{\rm BIC}_{\rm 900{\times}Lor}$ & 71 & 5720 & 16 \\
\hline
\end{tabular}
\end{table}

\subsection{$\eta$--$\gamma_{\rm Lor}$ correlation and the interpretation of large $\eta$}

Building on the $f\times{\rm Lor}$ crosschecking analysis in the previous subsection, this set of higher-$\gamma_{\rm Lor}$ variants also provides a direct illustration of the strong correlation between the Doppler-response parameters $\eta_i$ and the bulk Lorentz factors.
As discussed in  Appendix~\ref{apx:eta_i}, $\eta_i$ can be viewed as a compact, model-agnostic summary of the band-integrated Doppler response, i.e. the sensitivity of the band-limited intensity to small fractional changes in the inverse Doppler factor $\varepsilon$.

For a given geometric configuration (including the fitted angular offsets between the beam axes and the bulk-flow directions), mildly relativistic bulk motion produces a weaker Doppler contrast across rotational phase.
In such cases, the likelihood can compensate by driving $\eta_i$ to larger values, thereby increasing the sensitivity of the band-limited intensities to modest phase-dependent changes in $\varepsilon$ and reproducing a comparable effective Doppler imprint in the phaseograms.
Conversely, when $\gamma_{\rm Lor,1}$ and $\gamma_{\rm Lor,3}$ are forced to larger values while all other parameters are re-optimised, the required $\eta_i$ values decrease systematically (Table~\ref{tab:eta_vs_lor}).

Table~\ref{tab:eta_vs_lor} makes this trend explicit by comparing the baseline best-fitting $\eta_1$ and $\eta_3$ with two representative higher-$\gamma_{\rm Lor}$ variants ($f=3$ and $f=900$).
Even the moderate increase to $f=3$ already reduces the inferred $\eta$ values substantially in many bands (highlighted rows), while the more extreme $f=900$ variant further suppresses $\eta$ to values that are typically an order of magnitude smaller than in the baseline.
Importantly, these reductions occur despite allowing all other parameters to adjust freely, indicating that the large baseline $\eta$ values are not an isolated or fragile feature of a specific parametrisation, but rather reflect a robust degeneracy: the phaseogram likelihood constrains an \emph{effective} within-band Doppler response, which can be achieved either by (i) stronger Doppler contrasts from larger $\gamma_{\rm Lor}$ or (ii) a steeper within-band response encoded by larger $\eta$ when $\gamma_{\rm Lor}$ is only mildly relativistic.
This behaviour shows that apparently large $\eta_i$ values in the baseline fits can naturally accompany mildly relativistic $\gamma_{\rm Lor}$, rather than requiring extreme intrinsic spectral slopes.

\begin{table*}
\small
\centering
\caption{$\eta$--$\gamma_{\rm Lor}$ correlation illustrated by higher-Lorentz-factor variants.
For each pulsar and energy band we list the best-fitting Doppler-response parameters $\eta_{1}$ and $\eta_{3}$ for the baseline model, and the corresponding values obtained in restricted higher-$\gamma_{\rm Lor}$ variants in which both bulk Lorentz factors are fixed to $f$ times their baseline best-fitting values (here shown for $f=3$ and $f=900$), while all remaining parameters are refit freely (each variant has two fewer free parameters than the baseline).
Across the three pulsars, forcing larger $\gamma_{\rm Lor}$ systematically drives the best-fitting $\eta$ values downward, demonstrating a strong $\eta$--$\gamma_{\rm Lor}$ correlation: when the bulk motion is only mildly relativistic, the fit can compensate by adopting apparently large $\eta$ in order to reproduce a comparable band-integrated Doppler response in the phaseograms.}
\label{tab:eta_vs_lor}
\begin{tabular}{lcccccccccccc}
\hline
 & \multicolumn{4}{c}{Crab} & \multicolumn{4}{c}{Vela} & \multicolumn{4}{c}{Dragonfly} \\
 & 0.06 & 0.15 & 0.36 & 0.90
 & 0.06 & 0.15 & 0.36 & 0.90
 & 0.06 & 0.15 & 0.36 & 0.90 \\
$E_{\rm det}$ (GeV) & --   & --   & --   & --
 & --   & --   & --   & --
 & --   & --   & --   & --   \\
 & 0.15 & 0.36 & 0.90 & 3.0
 & 0.15 & 0.36 & 0.90 & 3.0
 & 0.15 & 0.36 & 0.90 & 3.0 \\
\hline
\multicolumn{13}{c}{Beams 1 and 2} \\
\hline
$\eta_{1,{\rm base}}$ &
  340.2 & 299.2 & 237.8 & 238.1 &
   13.2 &  17.5 &  21.2 &  28.1 &
  509.0 & 583.0 & 706.0 & 1049.0 \\
$\eta_{1,{\rm 3{\times}Lor}}$ &
  76.4 & 81.4 & 92.9 & 112.4 &
   15.5 &  14.5 &  16.2 &  18.8 &
  85.2 & 83.2 & 100.1 & 124.3 \\
$\eta_{1,{\rm 900{\times}Lor}}$ &
  74.8 & 79.8 & 91.1 & 110.3 &
   15.1 &  14.1 &  15.9 &  18.4 &
  83.7 & 80.7 & 96.5 & 117.2 \\
\hline
\multicolumn{13}{c}{Beams 3 and 4} \\
\hline
$\eta_{3,{\rm base}}$ &
  77.4 & 77.8 & 68.5 & 54.0 &
 478.0 &389.0 &369.0 &349.2 &
 532.0 &458.0 &321.2 &321.9 \\
$\eta_{3,{\rm 3{\times}Lor}}$ &
  17.8 & 17.5 & 15.9 & 13.6 &
 97.7 &173.6 &20.4 &18.0 &
 105.0 &18.1 &67.1 &178.3 \\
$\eta_{3,{\rm 900{\times}Lor}}$ &
  16.9 & 16.7 & 15.2 & 13.1 &
 74.6 &134.0 &16.4 &13.8 &
 241.9 &14.1 &79.1 &172.6 \\
\hline
\end{tabular}
\end{table*}

\subsection{Robustness of qualitative geometric inferences to an ultra-relativistic $f=900$ variant}

Building on the $f\times{\rm Lor}$ crosschecking analysis in an earlier subsection, we perform an ultra-relativistic stress test ($f=900$) to assess whether several qualitative geometric inferences are biased by potential mis-estimation of the bulk Lorentz factors. In this variant, $\gamma_{\rm Lor,1}$ and $\gamma_{\rm Lor,3}$	are fixed to $f$ times their baseline best-fitting values, while all remaining parameters are re-optimised.

As quantified by the BIC (Table~\ref{tab:BIC_highgamma}), the $f=900$ solutions are strongly disfavoured for high-S/N pulsars (particularly Vela). This is expected because, with abundant photon statistics, even modest structured residuals can accumulate into large likelihood penalties. The purpose of this subsection is therefore not to advocate the ultra-relativistic solutions, but to demonstrate that several \emph{structural} features of the inferred geometry persist even under such an extreme forcing of $\gamma_{\rm Lor}$ (Table~\ref{tab:geometry_LorX900}).

\begin{table}[t]
\small
\setlength{\tabcolsep}{2pt}
\centering
\caption{Geometric and kinematic parameters of the four-beam configurations for the ultra-relativistic $f=900$ variant.
Listed values are the best-fitting parameters obtained when $\gamma_{\rm Lor,1}$ and $\gamma_{\rm Lor,3}$
are fixed to $f=900$ times their baseline best-fitting values and all remaining parameters are re-optimised.
This table is provided for robustness assessment only: the $f=900$ variant is strongly disfavoured by the BIC
(Table~\ref{tab:BIC_highgamma}), but it retains the key qualitative features highlighted in the main text.}
\label{tab:geometry_LorX900}
\begin{tabular}{@{}lccc@{}}
\hline
     & Crab & Vela & Dragonfly \\
\hline
$\Theta_{\rm A}$                 ($^\circ$)        & $67.9$   & $54.5$   & $81.2$   \\
\hline
\multicolumn{4}{c}{Beams 1 and 2} \\
\hline
$d\sin\theta_{\rm B}/R_{\rm LC}$    & $0.93$  & $1.05$  & $0.89$  \\
$d\cos\theta_{\rm B}/R_{\rm LC}$    & $0.87$  & $0.37$  & $0.50$  \\
$t_0$                            (phase)           & $0.109$& $0.283$& $0.232$\\
$\theta_{\rm M}$                 ($^\circ$)        & $171.7$  & $147.3$  & $160.1$   \\
$\theta_{\rm N}$                 ($^\circ$)        & $96.2$   & $88.2$   & $81.1$   \\
$\theta_{\rm C}$                 ($^\circ$)        & $167.7$  & $125.2$  & $156.8$   \\
$\theta_{\rm Q}$                 ($^\circ$)        & $91.9$   & $92.9$   & $83.3$   \\
\hline
\multicolumn{4}{c}{Beams 3 and 4} \\
\hline
$D\sin\phi_{\rm B}/R_{\rm LC}$    & $0.98$  & $1.00$  & $0.85$  \\
$D\cos\phi_{\rm B}/R_{\rm LC}$    & $0.98$  & $0.93$  & $1.07$  \\
$T_0$                            (phase)           & $0.112$& $0.400$& $0.227$\\
$\phi_{\rm M}$                   ($^\circ$)        & $28.6$   & $27.4$   & $22.3$  \\
$\phi_{\rm N}$                   ($^\circ$)        & $2.4$   & $59.3$   & $3.1$   \\
$\phi_{\rm C}$                   ($^\circ$)        & $27.7$   & $32.9$   & $23.0$  \\
$\phi_{\rm Q}$                   ($^\circ$)        & $20.4$   & $59.7$   & $4.4$   \\
\hline
\multicolumn{4}{c}{Additional geometrical properties$^{\mathrm{a}}$} \\
\hline
$\Delta\theta$$^{\mathrm{b}}$ ($^{\circ}$)    &     4.1 & 22.3 & 3.4      \\
$\Delta\phi$$^{\mathrm{b}}$ ($^{\circ}$)     &     8.5 & 5.5 & 0.9        \\
$\Delta\varphi_{\rm site}$$^{\mathrm{c}}$ ($^{\circ}$)   & $1.3$ & $21.0$ & $-9.2$ \\
$\Delta\varphi_{\rm beam}$$^{\mathrm{c}}$ ($^{\circ}$)   & $72.9$ & $54.2$ & $69.7$ \\
\hline
\end{tabular}

\footnotesize{\textit{Notes.} $^{\mathrm{a}}$ These additional properties are computed by substituting some iterated parameters into specific equations. \\
$^{\mathrm{b}}$ Angular separation between a beam axis and a bulk flow direction ($\Delta\theta$ for Beams 1 and 2, $\Delta\phi$ for Beams 3 and 4). These are computed using the scalar product formula. \\
$^{\mathrm{c}}$ Azimuthal separation between the lower-altitude and higher-altitude emission sites ($\Delta\varphi_{\rm site}$), and azimuthal separation between the corresponding beam axes ($\Delta\varphi_{\rm beam}$). Relevant formulae are shown in another Appendix~\ref{apx:az_sep}. }
\end{table}

Specifically, the following qualitative inferences remain present in both the baseline solutions and the $f=900$ variants:
\begin{itemize}
\item \textbf{$\Theta_{\rm A}$--$\zeta$ agreement.}
The inferred $\Theta_{\rm A}$ remains broadly consistent with the independent viewing-angle proxy $\zeta$
inferred from PWN torus measurements.
\item \textbf{Altitude stratification and age ordering.}
Using $z_{\rm low}=d\cos\theta_{\rm B}$ and $z_{\rm high}=D\cos\phi_{\rm B}$ as height proxies, the $f=900$ variant still yields
$z_{\rm high}>z_{\rm low}$ in all three pulsars, with the vertical separation increasing from
$\Delta z\simeq0.11\,R_{\rm LC}$ (Crab; youngest in our sample) to $\Delta z\simeq0.56$--$0.57\,R_{\rm LC}$ (Vela and Dragonfly), preserving the age-ordered strengthening of altitude stratification.
\item \textbf{Outward radial component at high altitude.} Beams~3--4 and their corresponding bulk-flow directions continue to develop a radially outward component in the inferred geometry.
\item \textbf{Beam--motion near alignment.} The near alignment between the fitted beam axes and bulk-flow directions persists, indicating that the data continue to favour only small-to-moderate beam--motion offsets.
\item \textbf{Near co-longitudinality of the two sites per hemisphere.} The two emission sites within each hemisphere remain nearly co-longitudinal, i.e. located in a similar azimuthal sector.
\end{itemize}

In this context it is worth noting that two secondary interpretive remarks in the main text rely on the
\emph{BIC-favoured, mildly relativistic} solutions (Table~\ref{tab:geometry}) rather than being purely geometry-only invariants.
First, the inference that the higher-altitude outflows are gently focused toward the spin axis, motivated by
$d\sin\theta_{\rm B}>D\sin\phi_{\rm B}$ in Table~\ref{tab:geometry}, is not supported in the ultra-relativistic stress-test variant
(Table~\ref{tab:geometry_LorX900}).
Second, the general decline of the zenith angle of the lower-altitude bulk flow ($\theta_{\rm M}$) with age discussed in the main text is likewise not preserved
when $\gamma_{\rm Lor}$ is forced to extremely high values.
We therefore regard these two points as \emph{conditional interpretations} applicable within the statistically
preferred mildly relativistic regime, whereas the robustness exercise here is intended to highlight the
persistence of the more structural features listed above.

\subsection{Variants with $\varepsilon$-dependent beam widths}

Building on the baseline analysis with ``locally uniform'' beam shapes in the main text, we introduce variants with
$\varepsilon$-dependent beam widths to diagnose a specific potential degeneracy: the extent to which the fitted
Doppler-response indices $\eta_i$ in the benchmark model may be absorbing an unmodelled dependence of the
plasma-frame beam width $\Psi_c$ on the inverse Doppler factor $\varepsilon$.
This variant is introduced solely as a diagnostic and is \emph{not} treated as a competing benchmark model.
Specifically, for each pulsar and each $E_{\rm det}$ band, we allow the GND width of each beam pair to vary with
$\varepsilon$ as a power law,
\begin{equation}
\Psi_{c,1}(\varepsilon)=\Psi_{c,1}^{\rm norm}\,\varepsilon^{\kappa_1},
\qquad
\Psi_{c,3}(\varepsilon)=\Psi_{c,3}^{\rm norm}\,\varepsilon^{\kappa_3}.
\end{equation}

Allowing $\kappa_1$ and $\kappa_3$ to vary freely introduces strong covariance with the Doppler-response indices
$\eta_i$ (and partial covariance with $\beta$), leading to severe parameter degeneracies.
We therefore adopt a reduced diagnostic in which the width exponents are fixed to two small bracketing values,
$\kappa_1=\kappa_3=+0.05$ and $\kappa_1=\kappa_3=-0.05$, and refit all remaining parameters for each case.
This bracketing test is used only to quantify the sensitivity of $\eta$ to a controlled $\varepsilon$-dependence
of the beam width.

The resulting best-fitting $\eta$ values for $\kappa=0,\pm0.05$ are summarised in Table~\ref{tab:eta_kappa_fixed},
and the corresponding geometric/kinematic parameters are listed in Table~\ref{tab:geom_kappa_fixed}.
Together, these non-zero-$\kappa$ variants show that the fitted $\eta$ values can respond strongly even to a
mild imposed $\varepsilon$-dependence of the beam width, most prominently for Vela (and, to a lesser extent,
for Dragonfly, especially in $\eta_3$), indicating a substantial $\kappa$--$\eta$ covariance in the fits.
This strengthens the interpretation that the benchmark ($\kappa=0$) $\eta$ values may partly compensate for
an unmodelled $\varepsilon$-dependence of the beam width and of other beam-shape properties (including the
sharpness index $\beta$), as well as additional unresolved complexities of the intrinsic beam morphology.

\begin{table*}
\small
\centering
\caption{Best-fitting Doppler-response indices $\eta_1$ and $\eta_3$ for the non-zero-$\kappa$, $\varepsilon$-dependent beam-width variants.
For each pulsar and each $E_{\rm det}$ band we list the benchmark results ($\kappa=0$) and the refitted values obtained
when imposing an $\varepsilon$-dependent width $\Psi_{c,i}(\varepsilon)\propto\varepsilon^{\kappa_i}$ with
$\kappa_1=\kappa_3=+0.05$ and $\kappa_1=\kappa_3=-0.05$.
All other parameters are optimised freely in each case.}
\label{tab:eta_kappa_fixed}

\begin{tabular}{@{}lcccccccccccc@{}}
\hline
 & \multicolumn{4}{c}{Crab} & \multicolumn{4}{c}{Vela} & \multicolumn{4}{c}{Dragonfly} \\
 & 0.06 & 0.15 & 0.36 & 0.90
 & 0.06 & 0.15 & 0.36 & 0.90
 & 0.06 & 0.15 & 0.36 & 0.90 \\
$E_{\rm det}$ (GeV) & --   & --   & --   & --
 & --   & --   & --   & --
 & --   & --   & --   & --   \\
 & 0.15 & 0.36 & 0.90 & 3.0
 & 0.15 & 0.36 & 0.90 & 3.0
 & 0.15 & 0.36 & 0.90 & 3.0 \\
\hline
\multicolumn{13}{c}{Beams 1 and 2} \\
\hline
$\eta_{1}$ ($\kappa=0$)        &  340.2 & 299.2 & 237.8 & 238.1 &
   13.2 &  17.5 &  21.2 &  28.1 &
  509.0 & 583.0 & 706.0 & 1049.0 \\
$\eta_{1}$ ($\kappa=+0.05$)    &  330.5 & 290.9 & 228.1 & 227.1 & 10.9 & 10.3 & 12.4 & 18.2 & 543.0 & 596.0 & 720.0 & 1059.0  \\
$\eta_{1}$ ($\kappa=-0.05$)    &  345.1 & 304.2 & 242.3 & 241.5 & 19.0 & 18.6 & 17.7 & 22.2 & 458.0 & 516.0 & 630.0 & 958.0  \\
\hline
\multicolumn{13}{c}{Beams 3 and 4} \\
\hline
$\eta_{3}$ ($\kappa=0$)        &  77.4 & 77.8 & 68.5 & 54.0 &
 478.0 &389.0 &369.0 &349.2 &
 532.0 &458.0 &321.2 &321.9 \\
$\eta_{3}$ ($\kappa=+0.05$)    &  78.9 & 79.3 & 69.5 & 54.6 & 245.3 & 63.4 & 439.0 & 466.0 & 680.0 & 564.0 & 391.0 & 399.0  \\
$\eta_{3}$ ($\kappa=-0.05$)    &  71.8 & 72.2 & 63.8 & 50.4 & 1807.0 & 2214.0 & 1971.0 & 395.0 & 438.0 & 368.0 & 256.1 & 263.1  \\
\hline
\end{tabular}
\end{table*}

\begin{table*}
\small
\centering
\caption{Geometric and kinematic parameters for the non-zero-$\kappa$ beam-width variants.
Columns list the refitted best-fitting values for $\kappa_1=\kappa_3=+0.05$ and $\kappa_1=\kappa_3=-0.05$,
with all other parameters optimised freely.
This table is intended as a diagnostic comparison with the benchmark geometry rather than as a replacement of it.}
\label{tab:geom_kappa_fixed}

\begin{tabular}{@{}lcccccc@{}}
\hline
 & \multicolumn{3}{c}{$\kappa=+0.05$} & \multicolumn{3}{c}{$\kappa=-0.05$} \\
\cline{2-4}\cline{5-7}
 & Crab & Vela & Dragonfly & Crab & Vela & Dragonfly \\
\hline
$\Theta_{\rm A}$                 ($^\circ$)        & 54.8 & 58.3 & 83.9 & 56.2 & 68.9 & 84.3 \\
\hline
\multicolumn{7}{c}{Beams 1 and 2} \\
\hline
$d\sin\theta_{\rm B}/R_{\rm LC}$                    & 0.98 & 0.99 & 0.76 & 0.96 & 0.87 & 0.73 \\
$d\cos\theta_{\rm B}/R_{\rm LC}$                    & 0.49 & 0.41 & 0.77 & 0.51 & 0.61 & 0.84 \\
$t_0$                            (phase)           & 0.121 & 0.285 & 0.255 & 0.123 & 0.292 & 0.262 \\
$\gamma_{\mathrm{Lor},1}$                           & 1.12 & 72.30 & 1.03 & 1.12 & 56.10 & 1.03 \\
$\theta_{\rm M}$                 ($^\circ$)        & 165.8 & 148.0 & 102.8 & 165.8 & 137.0 & 100.8 \\
$\theta_{\rm N}$                 ($^\circ$)        & 87.1 & 86.2 & 93.2 & 87.6 & 91.4 & 94.9 \\
$\theta_{\rm C}$                 ($^\circ$)        & 161.9 & 121.1 & 101.3 & 161.9 & 114.9 & 99.0 \\
$\theta_{\rm Q}$                 ($^\circ$)        & 90.0 & 92.0 & 93.4 & 90.5 & 95.8 & 95.2 \\
\hline
\multicolumn{7}{c}{Beams 3 and 4} \\
\hline
$D\sin\phi_{\rm B}/R_{\rm LC}$                      & 0.93 & 0.64 & 0.64 & 0.92 & 0.48 & 0.62 \\
$D\cos\phi_{\rm B}/R_{\rm LC}$                      & 0.72 & 1.51 & 1.71 & 0.75 & 2.56 & 1.81 \\
$T_0$                            (phase)           & 0.137 & 0.400 & 0.305 & 0.137 & 0.430 & 0.299 \\
$\gamma_{\mathrm{Lor},3}$                           & 1.02 & 1.30 & 1.09 & 1.03 & 1.10 & 1.15 \\
$\phi_{\rm M}$                   ($^\circ$)        & 35.0 & 27.7 & 117.0 & 34.9 & 14.5 & 114.6 \\
$\phi_{\rm N}$                   ($^\circ$)        & 9.1 & 41.9 & 34.5 & 9.2 & 42.6 & 29.4 \\
$\phi_{\rm C}$                   ($^\circ$)        & 36.6 & 29.9 & 119.6 & 36.6 & 17.5 & 118.4 \\
$\phi_{\rm Q}$                   ($^\circ$)        & 27.2 & 42.1 & 34.9 & 26.7 & 42.8 & 29.9 \\
\hline
\multicolumn{7}{c}{Additional geometrical properties$^{\mathrm{a}}$} \\
\hline
$\Delta\theta$$^{\mathrm{b}}$ ($^{\circ}$)         & 4.0 & 27.2 & 1.5 & 4.0 & 22.4 & 1.8 \\
$\Delta\phi$$^{\mathrm{b}}$ ($^{\circ}$)           & 10.6 & 2.2 & 2.6 & 10.3 & 3.0 & 3.8 \\
$\Delta\varphi_{\rm site}$$^{\mathrm{c}}$ ($^{\circ}$)  & $-4.5$ & $-8.5$ & 5.5 & $-4.4$ & $-11.4$ & 1.7 \\
$\Delta\varphi_{\rm beam}$$^{\mathrm{c}}$ ($^{\circ}$)  & 58.4 & 41.4 & 64.0 & 59.4 & 41.6 & 67.1 \\
\hline
\end{tabular}

\footnotesize{\textit{Notes.} $^{\mathrm{a}}$ These additional properties are computed by substituting some iterated parameters into specific equations. \\
$^{\mathrm{b}}$ Angular separation between a beam axis and a bulk flow direction ($\Delta\theta$ for Beams 1 and 2, $\Delta\phi$ for Beams 3 and 4). These are computed using the scalar product formula. \\
$^{\mathrm{c}}$ Azimuthal separation between the lower-altitude and higher-altitude emission sites ($\Delta\varphi_{\rm site}$), and azimuthal separation between the corresponding beam axes ($\Delta\varphi_{\rm beam}$). Relevant formulae are shown in another Appendix~\ref{apx:az_sep}. }
\end{table*}

Specifically, the following qualitative inferences remain present in both the benchmark solutions and the
non-zero-$\kappa$ variants with $\varepsilon$-dependent beam widths:
(i) broad consistency between $\Theta_{\rm A}$ and the independently inferred $\zeta$,
(ii) a clear altitude stratification ($z_{\rm high}>z_{\rm low}$) in all three pulsars, with a substantially
larger vertical separation in adolescent Vela and Dragonfly than in very young Crab,
(iii) a mild to moderate collimation signature in the sense that $d\sin\theta_{\rm B}>D\sin\phi_{\rm B}$,
(iv) the general decline of the zenith angle of the lower-altitude bulk flow ($\theta_{\rm M}$) with age,
(v) an outward radial bulk-flow component at high altitude,
(vi) small-to-modest beam--motion misalignments, and
(vii) near co-longitudinality of the two emission sites within each hemisphere
($|\Delta\varphi_{\rm site}|\lesssim12^\circ$).
In addition, these cross-checking fits reinforce the inference that highly relativistic pulse-emitting plasma
is required only for Vela, whereas the pulse-emitting plasma of the Crab and Dragonfly remains mildly
relativistic.

\section{Systematic uncertainties from energy-band choices}\label{apx:band_edge}

The geometric parameters inferred from phaseograms can, in principle, depend on the precise choice of energy range used to construct the profiles. To quantify this effect we perform a series of cross-checking fits in which we vary the upper energy bound of the global analysis while keeping the lower bound fixed at 60~MeV. In addition to our baseline choice of $E_{\max}=3$~GeV, we repeat the entire four-beam fitting procedure for
$E_{\max} = 2.4,\ 2.7,\ 3.3,\ 3.6~\mathrm{GeV}$.
For each alternative $E_{\max}$ we re-distribute the four energy bands so that they still span the full 60~MeV--$E_{\max}$ interval with comparable photon statistics, then reconstruct the phaseograms and re-fit all three pulsars using the same methodology and initialisation strategy.

For a given geometric parameter $p$ (for example, $\Theta_A$ or $d\sin\theta_B$), we define the systematic uncertainty due to the high-energy cut as
$\sigma_{\rm sys}(p)
= \left\langle\,\bigl|p(E_{\max}) - p(3~\mathrm{GeV})\bigr|\,\right\rangle$,
where the average is taken over the four alternative $E_{\max}$ values. The total uncertainty quoted in the main text and in Table~\ref{tab:geometry} is the quadratic sum of the statistical and systematic components,
$\sigma_{\rm tot}^2(p) = \sigma_{\rm stat}^2(p) + \sigma_{\rm sys}^2(p)$.

We find that the systematic shifts are generally comparable to or smaller than the formal statistical errors for Vela and the Dragonfly pulsar, and somewhat larger for the Crab pulsar, but in all cases they do not alter the qualitative configuration of the four-beam geometry discussed in the main text.

\section{Additional robustness checks}

In addition to the $E_{\max}$ variations described above, we have performed a series of robustness tests. For each pulsar we verified that the inferred geometry is stable against changes in the phase-binning resolution and against modest variations of the region-of-interest radius, which probe the impact of background contamination at low energies. We also checked that small perturbations to the timing solutions (within their quoted uncertainties) do not change the resulting beam configuration beyond the combined statistical and systematic error bars.

Taken together, these tests indicate that the four-beam solutions reported in this work are not artefacts of a particular choice of binning or energy selection, but robust geometrical features required by the observed 60~MeV--3~GeV phaseograms of the Crab, Vela and Dragonfly pulsars.


\bibliography{Ref}{}
\bibliographystyle{aasjournalv7}



\end{document}